\documentclass[12pt]{article}
\usepackage{amssymb,amsmath,epsfig}
\allowdisplaybreaks

\begin{document}
\title{\bf Decoupled Charged Anisotropic Spherical Solutions in Rastall Gravity}
\author{M. Sharif$^1$ \thanks{msharif.math@pu.edu.pk}~
and M. Sallah$^{1,2}$ \thanks{malick.sallah@utg.edu.gm} \\
$^1$ Department of Mathematics and Statistics, The University of Lahore\\
1-KM Defence Road Lahore-54000, Pakistan.\\
$^2$ Department of Mathematics, The University of The Gambia,\\
Serrekunda, P.O. Box 3530, The Gambia.}
\date{}
\maketitle

\begin{abstract}
This paper uses the gravitational decoupling through the minimal
geometric deformation approach and extends a known isotropic
solution for a self-gravitating interior to two types of anisotropic
spherical solutions in Rastall gravity in the presence of
electromagnetic field. By deforming only the radial metric
component, the field equations are decoupled into two sets, the
first of which corresponds to an isotropic distribution of matter
while the second set contains the anisotropic source. We obtain a
solution of the first set by employing the charged isotropic
Finch-Skea ansatz, whereas a solution for the second set is obtained
by adopting two mimic constraints on the pressure and density. The
matching conditions at the stellar surface are explored with the
exterior geometry given by the deformed Reissner-Nordstrom
spacetime. For the two fixed values of the Rastall and charge
parameters, we investigate physical features of both solutions
through graphical analysis of the energy conditions, equation of
state parameters, surface redshift and compactness function. The
stability of both solutions is also studied through the Herrera
cracking approach and causality condition. We deduce that while both
solutions are physically viable, only the solution corresponding to
the pressure-like constraint is stable.
\end{abstract}

\section{Introduction}

General Relativity (GR) is regarded as one of the most important
pillars for comprehending the underlying notions of gravitational
phenomena and cosmology. Several cosmological measurements indicate
that celestial objects are not randomly distributed in the universe
but are instead systematically ordered. The study of such
interstellar objects and their physical attributes encourage us to
figure out cosmic expansion. This expansion is thought to be carried
out by an obscure type of energy known as dark energy which has a
repulsive tendency. Through the introduction of the cosmological
constant, an attempt has been made in GR to explain the nature and
gravitational effects of dark energy. However, several researchers
have argued the limitations/deficiency of the cosmological constant
in explaining the true dynamics of dark energy. To this end,
modified gravity theories were proposed. One such modified theory is
the Rastall gravity theory, proposed by Rastall in 1972 \cite{1}.
According to this theory, the stress-energy tensor which exhibits
null divergence in flat spacetime is not necessarily conserved in
curved spacetime geometry. Rastall gravity differs from GR in that
it includes the Ricci scalar through the Rastall parameter. Despite
being manually introduced, this factor alters not only the field
equations but also the way material fields are linked to the
gravitational interaction.

The minimal coupling concept clearly does not apply in this theory.
This, however, brings with it new and interesting ideas that may
help us to understand a variety of well studied phenomena, including
cosmological challenges, stellar systems, collapsed structures such
as black holes, gravitational waves and so on. Rastall gravity is
thus competitive with other modified theories of gravity such as
$f(\mathcal{R})$ and  $f(\mathcal{R},T)$ theories, where
$\mathcal{R}$ and $T$ are the Ricci scalar and trace of the
stress-energy tensor, respectively. An important feature of the
Rastall theory is that from the geometric perspective, any perfect
fluid solution of the Einstein field equations is also a solution of
the Rastall field equations. Furthermore, both GR and Rastall
gravity have the same vacuum solution with respect to black holes.
The Rastall field equations do not have an accompanying Lagrangian
density from which they can be calculated, but nevertheless,
generalize the GR field equations whilst maintaining the general
coordinate transformation of the theory.

The complex nature of the self-gravitating bodies is made clear by
the analytical solutions to the field equations. However, because
the field equations are non-linear, it is frequently challenging to
find solutions. A recently developed method to find workable
solutions is the gravitational decoupling by minimal geometric
deformation (MGD). In this method, the system of field equations is
divided into two sets by means of a linear transformation that
distorts the radial function of the line element. The extra source
is represented by the second set, while the first set describes the
seed sector. These two sets are treated separately, and the
superposition principle is used to get the solution for the entire
system. The MGD has its origins in the brane-world theory \cite{2}
and in \cite{3}, which were then expanded upon to look into new
black hole solutions \cite{4,5}. An exact interior solution for
isotropic spherically symmetric compact distributions was created by
Ovalle and Linares \cite{6}. This solution is essentially a
brane-world adaption of Tolman's solution.

In order to modify the temporal and radial metric functions of
spherical self-gravitating systems with a naked singularity at the
Schwarzschild radius, Casadio et al. \cite{7} reported a unique
external solution. Using the same method, Ovalle \cite{8} went on to
construct anisotropic solutions from ideal fluid configurations with
spherical symmetry. Isotropic interior solutions were extended by
Ovalle et al. \cite{9} to account for anisotropy. In an effort to
develop anisotropic solutions from known isotropic domains in
Rastall gravity and other modified theories, Maurya and his
associates \cite{10}-\cite{15} worked on the MGD technique. From a
known isotropic solution, Sharif and Saba \cite{16}-\cite{18}
developed charged/uncharged gravitational decoupled anisotropic
solutions and assessed the stability and sustainability of the
resultant solutions within the context of $f(G)$ gravity ($G$ is the
Gauss-Bonnet invariant). In order to verify the viability of compact
stars in the framework of various modified theories, numerous
researchers obtained anisotropic versions of the isotropic source
\cite{19}-\cite{22}.

The participation of the electromagnetic field in celestial
formations has an intriguing effect on how their evolution can be
researched and examined. The impact of charge on stellar bodies in
GR has been extensively studied through a vast corpus of literature
and updated models. A static sphere that constitutes a charged
perfect fluid was discussed by Xingxiang \cite{23}. By matching the
outer Reissner-Nordstrom (RN) metric with the inner geometry, Das et
al. \cite{24} investigated charged static spherical solutions. The
field equations for a shearfree charged object were numerically
solved by Sharif and Bhatti \cite{25}, who also verified the
solution's feasibility using the energy conditions. Murad \cite{26}
investigated anisotropic charged celestial objects with the
assumption of a particular metric potential. In the presence of an
electric field, many features characterizing the internal
composition of self-gravitating entities have been investigated
\cite{27}-\cite{29}. Using the Krori-Barua spacetime and the MGD
method, Sharif and Sadiq \cite{30} investigated the effect of charge
and created two anisotropic solutions. In
$f(\mathcal{R},T,R_{\gamma\upsilon}T^{\gamma\upsilon})$ gravity,
Sharif and Naseer \cite{31,32} obtained charged as well as uncharged
anisotropic spherical solutions by minimally deforming the
Krori-Barua ansatz. Sharif and Hassan \cite{33,34} also obtained
charged-uncharged anisotropic spherical solutions in the context of
$f(G,T)$ gravity using the MGD approach.

In addition to providing insights into useful astrophysical
phenomena, some researchers \cite{35} were able to obtain precise
solutions by accounting for a certain sort of anisotropy. This
allowed them to demonstrate that spherical stars may sustain
positive and finite pressures and densities. Gleiser and Dev
\cite{36} demonstrated that anisotropy can sustain stars with a
specific compactness $\frac{M}{2R}=\frac{2}{9}$  ($M$ and $R$ denote
mass and radius, respectively) and concluded that stable
configurations exist for specific adiabatic index relative to
isotropic fluids. By attaining precise solutions for spherically
symmetric anisotropic matter distributions with a linear equation of
state (EoS), Sharma and Maharaj \cite{37} made significant
advancements in the modeling of compact stars. As part of his
investigation on the stability of self-gravitating models, Herrera
\cite{38} proposed the concepts of cracking and overturning to
examine the behavior of isotropic and anisotropic structures
following perturbations. His findings showed that perfect fluid
distributions are stable while anisotropic fluid distributions
fracture. By adding sound speed, Abreu et al. \cite{39} expanded the
concept of cracking to study anisotropic spherical structures. They
concluded that when the square of the tangential sound speed is
greater than the square of the radial sound speed, the system
becomes unstable.

This paper examines the relationship between Rastall gravity and
charged anisotropic spherical systems in an effort to create
solutions that appropriately capture the gravitational behavior of
self-gravitating systems. The rest of the paper is structured as
follows. Effective parameters are determined by analyzing the
Rastall field equations for a static spherically symmetric matter
distribution in section \textbf{2}. In section \textbf{3}, we split
the Rastall field equations into two workable sets using the MGD
technique. The matching conditions at the stellar surface are also
examined. In section \textbf{4}, by extending a known charged
perfect fluid ansatz, we obtain two solutions for anisotropic
spherical source. We examine physical properties spanning from the
viability to stability of our acquired solutions. Finally, section
\textbf{5} discusses a summary of our findings.

\section{Rastall Gravity Theory}

The Rastall gravity theory arose as a result of the rejection of the
fundamental premise that the stress-energy tensor freely diverges in
a curved spacetime. The Rastall field equations, given by
\begin{equation}\label{1}
R_{\gamma\varrho}-\frac{1}{2}\mathcal{R}g_{\gamma\varrho}=\kappa
(T^{(m)}_{\gamma\varrho}-\lambda \mathcal{R}g_{\gamma\varrho}),
\end{equation}
are consistent with the assumption that
\begin{equation}\label{2}
\nabla_\varrho T^{\gamma\varrho}=\lambda
g^{\gamma\varrho}\nabla_\varrho \mathcal{R},
\end{equation}
and reduce to the Einstein's field equations in the event
$\lambda=0$. The non-conservation of the stress-energy tensor
\eqref{2} as proposed by Rastall, induces a non-minimal coupling
between matter and geometry. In the preceding equations, $\kappa$ is
the coupling constant and $\lambda$ is the Rastall parameter that
causes the deviation from GR and by which the Ricci scalar is
non-minimally coupled into the theory. The term
$T^{(m)}_{\gamma\varrho}$ denotes a charged perfect fluid matter
configuration given by
\begin{equation}\label{3}
T^{(m)}_{\gamma\varrho}=(\rho+P)u_\gamma u_\varrho -
P\,g_{\gamma\varrho}+\frac{1}{4\pi}\big[\frac{1}{4}g_{\gamma\varrho}F^{\tau\upsilon}
F_{\tau\upsilon}-F^\tau_\gamma F_{\varrho\tau}\big],
\end{equation}
where $u^\gamma=e^{-\frac{\beta(r)}{2}}~\delta^\gamma_0$ is the
fluid 4-velocity while $\rho$ and $P$ represent the energy density
and isotropic pressure, respectively. The term
$F_{\gamma\varrho}=\alpha_{\varrho,\gamma}-\alpha_{\gamma,\varrho}$
denotes the Maxwell field tensor with
$\alpha_\varrho=\alpha(r)\delta^0_\varrho$ as 4-potential. This
Maxwell tensor satisfies the Maxwell field equations given by
\begin{equation}\label{4}
F^{\gamma\varrho}_{;\varrho}=4\pi J^\gamma,\quad
F_{[\gamma\varrho;\eta]},
\end{equation}
where $J^\gamma$ is the 4-current which can be written in terms of
charged density $\zeta$ as $J^\gamma=\zeta(r) u^\gamma$.

For the purpose of describing our interior geometry, we shall
consider a static spherically symmetric spacetime in
Schwarzschild-like coordinates as
\begin{equation}\label{5}
ds^2_{-}=e^{\beta(r)}dt^2-e^{\eta(r)}dr^2-r^2(d\theta^2+
\sin^2\theta d\phi^2),
\end{equation}
where the areal radius $r$ ranges from the stars center $(r=0)$ to
an arbitrary point $(r=R)$ on the surface of the star. The Maxwell
field equation for our spacetime becomes
\begin{equation}\label{6}
\alpha^{\prime\prime}+\bigg(\frac{2}{r}-\frac{\beta^\prime}{2}-
\frac{\eta^\prime}{2}\bigg)\alpha^\prime=4\pi\zeta
e^{\frac{\beta}{2}+\eta},
\end{equation}
where $^\prime=\frac{\partial}{\partial r}$. Upon integration of the
above equation, we have
\begin{equation}\label{7}
\alpha^\prime=\frac{e^{\frac{\beta+\eta}{2}}q(r)}{r^2},
\end{equation}
$q(r)=4\pi\int_0^r\zeta e^{\frac{\eta}{2}}r^2dr$ indicates the total
charge in the interior of the sphere. By defining
\begin{equation}\label{8}
\bar{T}^{(eff)}_{\gamma\varrho}=T^{(m)}_{\gamma\varrho}-\lambda
\mathcal{R}g_{\gamma\varrho},
\end{equation}
we can rewrite the field equations \eqref{1} as
\begin{equation}\label{9}
G_{\gamma\varrho}=\kappa\bar{T}_{\gamma\varrho}^{(eff)}.
\end{equation}
This shows that the original Rastall field equations can always be
restructured to remold the Einstein's field equations, thereby
regaining the standard result
$\nabla_\varrho\bar{T}^{\gamma\varrho(eff)}=0$. This rearrangement
can also be performed in other modified gravity theories such as
$f(\mathcal{R})$, $f(\mathcal{R},T)$ theories among others,
irrespective of the conservation of the stress-energy tensor.

Upon contracting the field equations \eqref{1}, we can write the
Ricci scalar as
\begin{equation}\label{10}
\mathcal{R}=\frac{\kappa T}{4\lambda\,\kappa-1},
\end{equation}
which can, in turn, be used to rewrite the effective stress-energy
tensor \eqref{8} as
\begin{equation}\label{11}
\bar{T}^{(eff)}_{\gamma\varrho}=T^{(m)}_{\gamma\varrho}-
\frac{\varepsilon\,T}{4\,\varepsilon -1}\,g_{\gamma\varrho},
\end{equation}
where $\varepsilon=\lambda\,\kappa$. We shall proceed with the
assumption that $\kappa=1$ so that $\varepsilon=\lambda$. At this
point, it is clear that $\lambda=\frac{1}{4}$ depicts a
non-realistic scenario and must therefore be avoided. The components
of the effective stress-energy tensor \eqref{11} are thus obtained
as
\begin{align}\label{12}
\bar{T}^{(eff)}_{00}&=g_{00}\bigg(\frac{3\lambda(\rho+P)-\rho}{4\lambda-1}\bigg)
+g_{00}\bigg(\frac{q^2}{8\pi r^4}\bigg),\\\label{13}
\bar{T}^{(eff)}_{11}&=-g_{11}\bigg(\frac{\lambda(\rho+P)-P}{4\lambda-1}\bigg)
+g_{11}\bigg(\frac{q^2}{8\pi r^4}\bigg),\\\label{14}
\bar{T}^{(eff)}_{22}&=-g_{22}\bigg(\frac{\lambda(\rho+P)-P}{4\lambda-1}\bigg)
-g_{22}\bigg(\frac{q^2}{8\pi r^4}\bigg).
\end{align}
We now shift our attention to the field equations for multiple
matter sources, given by
\begin{equation}\label{15}
R_{\gamma\varrho}-\frac{1}{2}\mathcal{R}g_{\gamma\varrho}=
T_{\gamma\varrho}^{(tot)},
\end{equation}
with
\begin{equation}\label{16}
T_{\gamma\varrho}^{(tot)}=\bar{T}^{(eff)}_{\gamma\varrho}
+\delta\,\Theta_{\gamma\varrho}.
\end{equation}
Here, $\bar{T}^{(eff)}_{\gamma\varrho}$ is the effective
energy-momentum tensor given by Eq.\eqref{8} and the term
$\Theta_{\gamma\varrho}$ is an additional source gravitationally
coupled to the seed source through the decoupling constant $\delta$,
that may generate anisotropy in self-gravitating fields.

New fields such as scalar, tensor and vector fields may well be
contained in the source $\Theta_{\gamma\varrho}$. By virtue of its
definition, the total energy-momentum tensor in \eqref{16} must now
satisfy the conservation equation given by
\begin{equation}\label{17}
T^{\gamma\,(tot)}_{\varrho~;~\gamma}=0.
\end{equation}
The corresponding Rastall field equations turn out to
\begin{eqnarray}\label{18}
e^{-\eta}\bigg(\frac{\eta^\prime}{r}-\frac{1}{r^2}\bigg)
+\frac{1}{r^2}=\frac{3\lambda(\rho+P)-\rho}{4\lambda-1}+\frac{q^2}{8\pi
r^4} +\delta\,\Theta^0_0,\\\label{19}
e^{-\eta}\bigg(\frac{\beta^\prime}{r}+\frac{1}{r^2}\bigg)
-\frac{1}{r^2}=\frac{\lambda(\rho+P)-P}{4\lambda-1}-\frac{q^2}{8\pi
r^4} - \delta\,\Theta^1_1, \\\label{20}
e^{-\eta}\bigg(\frac{\beta^{\prime\prime}}{2}+\frac{\beta^{\prime^2}}{4}
-\frac{\beta^\prime \eta^\prime}{4}+\frac{\beta^\prime -
\eta^\prime}{2r}\bigg)=\frac{\lambda(\rho+P)-P}{4\lambda-1}+\frac{q^2}{8\pi
r^4} - \delta\Theta^2_2.
\end{eqnarray}
With respect to the system \eqref{18} - \eqref{20}, the conservation
equation in \eqref{17} now reads
\begin{equation}\label{21}
\bar{P}^\prime(r)+\frac{\beta^\prime(r)}{2}(\bar{\rho}+\bar{P})
+\frac{2\delta}{r}(\Theta^2_2-\Theta^1_1)+\frac{\delta\beta^\prime(r)}{2}
(\Theta^0_0-\Theta^1_1)-\delta\bigg(\Theta^1_1(r)\bigg)^\prime-
\frac{qq^\prime}{4\pi r^4}=0,
\end{equation}
where $\bar{\rho}=\frac{3\lambda(\rho+P)-\rho}{4\lambda-1}$ and
$\bar{P}=\frac{\lambda(\rho+P)-P}{4\lambda-1}\,.$ Equations
\eqref{18}-\eqref{21} constitute four non-linear differential
equations with eight unknowns given by
$(\beta,\eta,\rho,P,q,\\\Theta^0_0,\Theta^1_1,\Theta^2_2).$ From
this system, we identify three effective matter components given by
\begin{align}\label{22}
\rho^{eff}=\rho + \delta\,\Theta^0_0,\quad
\bar{P}_r^{eff}=P-\delta\,\Theta^1_1,\quad \bar{P}_t^{eff} = P
-\delta\Theta^2_2.
\end{align}
These definitions of the effective parameters indicate that the
source $\Theta_{\gamma\varrho}$ can induce an anisotropy within the
stellar distribution given by
\begin{equation}\label{23}
\bar{\Delta}^{eff}=P_t^{(eff)}(r)-P_r^{(eff)}(r)=\delta\,(\Theta^1_1-\Theta^2_2).
\end{equation}
In what follows, we explore the MGD technique in a bid to solve the
field equations \eqref{18}-\eqref{20}.

\section{Decoupling with MGD}

In order to find a solution to the field equations
\eqref{18}-\eqref{20}, we apply the gravitational decoupling
procedure with the MGD approach. Via this approach, the system will
be modified in such a way that the field equations associated with
the extra source $\Theta_{\gamma\varrho}$ assume the form of the
effective quasi Einstein equations. To proceed, we assume a charged
perfect fluid solution $(\xi,\chi,\rho,P,q)$ described by the line
element
\begin{equation}\label{24}
ds^2=e^{\xi(r)}-\frac{dr^2}{\chi(r)}-r^2(d\theta^2+\sin^2\theta
d\phi^2),
\end{equation}
where $\chi(r)=1-\frac{2m}{r}+\frac{q^2}{r^2},~m$ denotes the
Misner-Sharp mass of the stellar object with perfect matter
configuration. To introduce the effects of the source
$\Theta_{\gamma\varrho}$ on charged isotropic solution, we consider
the following linear transformations on the metric components
\begin{equation}\label{25}
\xi(r)\mapsto\beta(r)=\xi(r),\quad \chi(r)\mapsto
e^{-\eta(r)}=\chi(r)+\delta g^\ast(r),
\end{equation}
where $g^\ast$ is the deformation associated with the radial
component of the metric function. It can be observed that the MGD
approach only acts on the radial component of the metric and leaves
the temporal metric component un-deformed. By substituting the
deformed radial coefficient in the field Eqs.\eqref{18}-\eqref{20},
two sets of differential equations are obtained. The first set
corresponds to a perfect fluid configuration with $\delta=0$ and
reads
\begin{align}\label{26}
\frac{1}{r^2}-\frac{\chi}{r^2}-\frac{\chi^\prime}{r}&=
\frac{3\lambda(\rho+P)-\rho}{4\lambda-1}+\frac{q^2}{8\pi
r^4},\\\label{27}
\chi\bigg(\frac{\beta^\prime}{r}+\frac{1}{r^2}\bigg)
-\frac{1}{r^2}&=\frac{\lambda(\rho+P)-P}{4\lambda-1}-\frac{q^2}{8\pi
r^4},\\\label{28}
\chi\bigg(\frac{\beta^{\prime\prime}}{2}+\frac{\beta^{\prime^2}}{4}
+\frac{\beta^\prime}{2r}\bigg)+\chi^\prime\bigg(\frac{\beta^\prime}{4}
+\frac{1}{2r}\bigg)&=\frac{\lambda(\rho+P)-P}{4\lambda-1}+\frac{q^2}{8\pi
r^4},
\end{align}
with the associated conservation equation
\begin{equation}\label{29}
\bar{P}^\prime(r)+\frac{\beta^\prime(r)}{2}\,(\bar{\rho}+\bar{P})
-\frac{qq^\prime}{4\pi r^4}=0.
\end{equation}

The second set of equations entails the source
$\Theta_{\gamma\varrho}$ given by
\begin{eqnarray}\label{30}
\Theta_0^0&=&-\frac{g^{\ast^\prime}}{r}-\frac{g^\ast}{r^2},
\\\label{31}
\Theta^1_1&=&-g^\ast\bigg(\frac{\beta^\prime}{r}+\frac{1}{r^2}\bigg),
\\\label{32}
\Theta^2_2&=&-g^\ast\bigg(\frac{\beta^{\prime\prime}}{2}
+\frac{\beta^{\prime^2}}{4}+\frac{\beta^\prime}{2r}
\bigg)-g^{\ast^\prime}\bigg(\frac{\beta^\prime}{4}+\frac{1}{2r}\bigg),
\end{eqnarray}
and satisfies the conservation equation
\begin{equation}\label{33}
\frac{2}{r}(\Theta^2_2-\Theta^1_1)+\frac{\alpha^\prime(r)}{2}(\Theta^0_0-\Theta^1_1)
-\bigg(\Theta^1_1(r)\bigg)^\prime=0.
\end{equation}
The system given by Eqs.\eqref{30}-\eqref{32} resembles the field
equations for the anisotropic spherically symmetric matter
configuration with effective parameters $\rho^{eff}=\Theta_0^0$,
$P_r^{eff}=-\Theta_1^1,~P_t^{eff}=-\Theta_2^2,$ and the
corresponding metric becomes
\begin{equation}\nonumber
ds^2=e^{\beta(r)}dt^2-\frac{dr^2}{g^\ast(r)}-r^2\big(d\theta^2+\sin^2\theta
d\phi^2\big).
\end{equation}
However, due to a missing $\frac{1}{r^2}$ term in the right hand
side of the first two equations, this system does not represent
typical field equations for anisotropic matter configuration. The
matter components thus become
\begin{align}\nonumber
\rho^{eff}+\frac{q^2}{8\pi
r^4}&=\Theta^{\ast\,0}_0=\Theta_0^0+\frac{1}{r^2},\\\nonumber
P_r^{eff}-\frac{q^2}{8\pi
r^4}&=\Theta^{\ast\,1}_1=\Theta_1^1+\frac{1}{r^2},\\\nonumber
P_t^{eff}+\frac{q^2}{8\pi
r^4}&=\Theta^{\ast\,2}_2=\Theta_2^2=\Theta^{\ast\,3}_3=\Theta_3^3.
\end{align}

We now shift our attention to the junction conditions which provide
the governing rules for the smooth matching of the interior and
exterior space-time geometries at the surface of the star (where
$r=R$). Our interior spacetime geometry is given by the deformed
metric
\begin{equation}\label{34}
ds^2_-=e^{\beta(r)}dt^2-\frac{1}{\left(1-\frac{2m(r)}{r}+\delta
g^\ast(r)+\frac{q^2}{r^2}\right)}dr^2-r^2(d\theta^2+\sin^2\theta
d\phi^2),
\end{equation}
which is to be matched with the general outer metric given by
\begin{equation}\label{35}
ds^2_+=e^{\beta(r)}dt^2-e^{\eta(r)}dr^2-r^2(d\theta^2+\sin^2\theta
d\phi^2).
\end{equation}
Hence, the continuity of the first fundamental form
$\left([ds^2]_{\Sigma}=0\right)$ of junction conditions at the
hypersurface $\Sigma$ yields
\begin{equation}\label{36}
\beta(R)_-=\beta(R)_+,
\end{equation}
and
\begin{equation}\label{37}
1-\frac{2\,M_0}{R}+\frac{Q_0^2}{R^2}+\delta g^\ast_R=e^{-\eta(R)_+},
\end{equation}
where $\chi=e^{-\eta}-\delta g^\ast$. Also, $M_0=m(R),~Q_0^2=q^2(R)$
and $g^\ast_R$ denote the mass, charge and deformation at the
surface of the star.

Similarly, the continuity of the second fundamental form
($[T_{\gamma\varrho},S^\varrho]_{\Sigma}=0,~S^\varrho$ denotes a
unit 4-vector) gives
\begin{equation}\label{38}
P(R)-\frac{Q_0^2}{8\pi R^4}-\delta\left(\Theta^1_1(R)\right)_-
=-\delta\left(\Theta^1_1(R)\right)_+.
\end{equation}
Substituting Eq.\eqref{31} for the interior geometry in \eqref{38}
yields
\begin{equation}\label{39}
P(R)-\frac{Q_0^2}{8\pi R^4}+\delta
g^\ast(R)\bigg(\frac{\beta^\prime(R)}{R}+\frac{1}{R^2}\bigg)
=-\delta\left(\Theta^1_1(R)\right)_+.
\end{equation}
Using Eq.\eqref{31} for the outer geometry in \eqref{39}, we obtain
\begin{equation}\label{40}
P(R)-\frac{Q_0^2}{8\pi R^4}+\delta
g^\ast(R)\bigg(\frac{\beta^\prime(R)}{R}+\frac{1}{R^2}\bigg) =\delta
b^\ast(R)\,\bigg[\frac{1}{R^2}+\frac{2\mathcal{M}R-2\mathcal{Q}^2}
{R^2(R^2-2\mathcal{M}R+\mathcal{Q}^2)}\bigg],
\end{equation}
where $\mathcal{M}$ and $\mathcal{Q}$ denote the mass and charge in
the exterior region and $b^\ast(R)$ indicates the deformation on the
outer RN solution by the source $\Theta_{\gamma\varrho}$, as shown
below
\begin{equation}\label{41}
ds^2=\left(1-\frac{2\mathcal{M}}{r}+\frac{\mathcal{Q}^2}{r^2}\right)dt^2
-\frac{1}{\left(1-\frac{2\mathcal{M}}{r}+\frac{\mathcal{Q}^2}{r^2}
+\delta~b^\ast(r)\right)}dr^2-r^2 d\Omega^2.
\end{equation}
Equations \eqref{36}, \eqref{37} and \eqref{40} are the necessary
and sufficient conditions for the smooth matching of the deformed
interior metric \eqref{34} and the deformed RN metric \eqref{41}.

\section{Anisotropic Solutions}

Here, we obtain charged anisotropic spherical solutions for compact
stellar configuration by assuming a charged isotropic seed solution
of the field equations \eqref{18}-\eqref{20}. For this purpose, we
adopt the charged Finch-Skea solution due to its non-singularity and
physical plausibility. The solution is given by \cite{40}
\begin{eqnarray}\label{42}
e^{\beta(r)}&=&\bigg[A+\frac{1}{2}B r\sqrt{Cr^2}\bigg]^2,
\\\label{43}
\chi(r)&=&\frac{1}{1+Cr^2},\\\nonumber \rho&=&\frac{-C}{2 \sqrt{C
r^2}\left(C r^2+1\right)^2 \left(2 A+B r \sqrt{C
r^2}\right)^2}\bigg[4 A^2 \sqrt{C r^2} \bigg(C (4 \lambda -1)
r^2\\\nonumber &+&12 \lambda-6\bigg)-2 B r \ln\left(A+\frac{1}{2} B
r \sqrt{C r^2}\right) \bigg(2 B (4 \lambda -1) r \sqrt{C r^2}
\left(C r^2+1\right)\\\nonumber &\times& \ln\left(A+\frac{1}{2} B r
\sqrt{C r^2}\right)+2 A \left(C (8 \lambda +1) r^2+12 \lambda
\right)+B r \sqrt{C r^2}\\\nonumber &\times& \left(3 C r^2+4 \lambda
+2\right)\bigg)+4 A B C r^3 \left(C (4 \lambda -1) r^2+12 \lambda
-6\right)\\\label{44} &+&B^2 r^2 \sqrt{C r^2} \left(C r^2 \left(C (4
\lambda -1) r^2-4 \lambda -2\right)-16 \lambda
+4\right)\bigg],\\\nonumber P&=&\frac{C}{2 \sqrt{C r^2} \left(C
r^2+1\right)^2 \left(2 A+B r \sqrt{C r^2}\right)^2} \bigg[4 A^2
\sqrt{C r^2} \bigg(C (4 \lambda -1) r^2\\\nonumber &+&12 \lambda
-2\bigg)+2 B r \ln \left(A+\frac{1}{2} B r \sqrt{C r^2}\right)
\bigg(-2 B (4 \lambda -1) r \sqrt{C r^2}\\\nonumber &\times&\left(C
r^2+1\right)\ln \left(A+\frac{1}{2} B r \sqrt{C r^2}\right)+A
\left(2 C (3-8 \lambda ) r^2-24 \lambda +8\right)\\\nonumber &+&B r
\sqrt{C r^2} \left(C r^2-4 \lambda +2\right)\bigg)+4 A B C r^3
\left(C (4 \lambda -1) r^2+12 \lambda -2\right)\\\label{45} &+&B^2
r^2 \sqrt{C r^2} \left(C r^2 \left(C (4 \lambda -1) r^2-4 \lambda
+2\right)-16 \lambda +4\right)\bigg],\\\nonumber q^2&=&\frac{4 \pi C
r^5}{\left(C r^2+1\right)^2 \left(2 A+B r \sqrt{C
r^2}\right)^2}\bigg[r \bigg(4 A^2 C+4 A B C r \sqrt{C
r^2}+B^2\\\nonumber &\times& \left(C r^2+2\right)^2\bigg)+2 B \ln
\left(A+\frac{1}{2} B r \sqrt{C r^2}\right) \bigg(4 B r \left(C
r^2+1\right)\\\label{46} &\times& \ln\left(A+\frac{1}{2} B r \sqrt{C
r^2}\right)-2 A \sqrt{C r^2}-B r \left(3 C r^2+2\right)\bigg)\bigg],
\end{eqnarray}
where the constants $A,~B$ and $C$ can be determined from the
matching conditions. With the RN metric as our exterior spacetime,
the matching conditions yield
\begin{eqnarray}\nonumber
A&=&\sqrt{\frac{R^2-2M_0R+Q_0^2}{R^2}}-\frac{M_0R-Q_0^2}
{2R\sqrt{R^2-2M_0R+Q_0^2}},\\\label{47}
B&=&\frac{MR-Q_0^2}{R^2\sqrt{2M_0R-Q_0^2}},\quad
C=\frac{1}{R^2-2M_0R+Q_0^2}-\frac{1}{R^2},
\end{eqnarray}
with the compactness $\frac{M_0}{2R}<\frac{2}{9}$. These values
ensure the surface continuity of the interior and exterior
geometries and will most certainly be altered upon addition of the
source $\Theta_{\gamma\varrho}$. We now find anisotropic solutions
by setting $\delta\neq 0$ in the interior geometry and utilize
Eqs.\eqref{42} and \eqref{43} as our temporal and radial metric
coefficients, respectively. The deformation function $g^\ast(r)$ is
related to the source $\Theta_{\gamma\varrho}$ through
Eqs.\eqref{30}-\eqref{32}, which is a system of three equations in
four unknowns. It thus suffices to impose a single constraint to
close this system. In what follows, we obtain two anisotropic
solutions by imposing a single constraint in each case.

\subsection{Solution I}

We impose a constraint on $\Theta_1^1$ and obtain a solution of the
field Eqs.\eqref{30}-\eqref{32} for $g^\ast$ and
$\Theta_{\gamma\varrho}$. The interior geometry is compatible with
the exterior spacetime (given by the RN metric) whenever
$P(R)-\frac{Q_0^2}{8\pi
R^4}\sim\delta\left(\Theta_1^1(R)\right)_{-},$ leading to the
simplest choice
\begin{equation}\label{48}
P-\frac{q^2}{8\pi r^4}=\Theta_1^1~.
\end{equation}
From the equation above, we obtain an explicit expression for the
deformation function given by
\begin{align}\nonumber
g^\ast(r)&=\frac{-Cr^2}{\left(C r^2+1\right)^2 \big[BCr^3 \left(4
\ln\left(A+\frac{1}{2} B r \sqrt{C r^2}\right)+1\right)+2 A \sqrt{C
r^2}~\big]}\\\nonumber &\times\frac{1}{\left(2 A+B r \sqrt{C
r^2}\right)}\bigg[4 A^2 \sqrt{C r^2} \left(C (2 \lambda -1) r^2+6
\lambda -1\right)-2 B r\\\nonumber &\times \ln\left(A+\frac{1}{2} B
r \sqrt{C r^2}\right) \bigg(B (4 \lambda +1) r \sqrt{C r^2} \left(C
r^2+1\right)\\\nonumber &\times \ln\left(A+\frac{1}{2} B r \sqrt{C
r^2}\right)+4 A \left(C (2 \lambda -1) r^2+3 \lambda
-1\right)\\\nonumber &+2 B r \sqrt{C r^2} \left(-C r^2+\lambda
-1\right)\bigg)+4 A B C r^3 \left(C (2 \lambda -1) r^2+6 \lambda
-1\right)\\\label{49} &+B^2 r^2 \sqrt{C r^2} \left(C r^2 \left(C (2
\lambda -1) r^2-2 \lambda -1\right)-8 \lambda \right)\bigg].
\end{align}
Using the above equation, we can obtain an expression for the radial
metric function defined in Eq.\eqref{25} as
\begin{align}\nonumber
e^{-\eta(r)}&=\frac{-C \delta  r^2}{\left(C r^2+1\right)^2 \left(BC
r^3 \left(4 \ln \left(A+\frac{1}{2} B r \sqrt{C
r^2}\right)+1\right)+2 A \sqrt{C r^2}\right)}\\\nonumber
&\times\frac{1}{\left(2 A+B r \sqrt{C r^2}\right)} \bigg[4 A^2
\sqrt{C r^2} \left(C (2 \lambda -1) r^2+6 \lambda -1\right)-2 B
r\\\nonumber &\times \ln\left(A+\frac{1}{2} B r \sqrt{C r^2}\right)
\bigg(B (4 \lambda +1) r \sqrt{C r^2} \left(C r^2+1\right) \ln
\bigg(A\\\nonumber &+\frac{1}{2} B r \sqrt{C r^2}\bigg)+4 A \left(C
(2 \lambda -1) r^2+3 \lambda -1\right)+2 B r \sqrt{C r^2}
\bigg(\lambda -1\\\nonumber &-Cr^2\bigg)\bigg)+4 A B C r^3 \left(C
(2 \lambda -1) r^2+6 \lambda -1\right)+B^2 r^2 \sqrt{C r^2} \bigg(C
r^2\\\label{50}&\times \left(C (2 \lambda -1) r^2-2 \lambda
-1\right)-8 \lambda \bigg)\bigg]+\frac{1}{C r^2+1}~.
\end{align}
The interior metric functions \eqref{42} and \eqref{50} denote the
minimally deformed Finch-Skea solution. The expressions for the
effective parameters together with the induced anisotropy are given
in the Appendix.
\begin{figure}\center
\epsfig{file=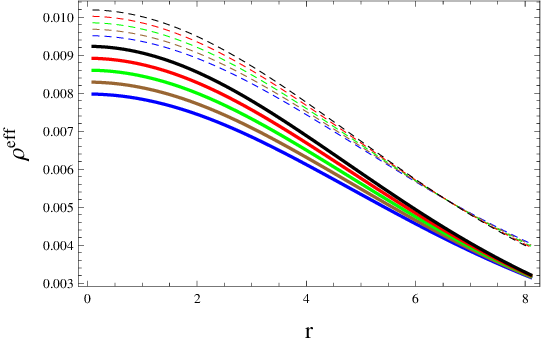,width=0.475\linewidth}
\epsfig{file=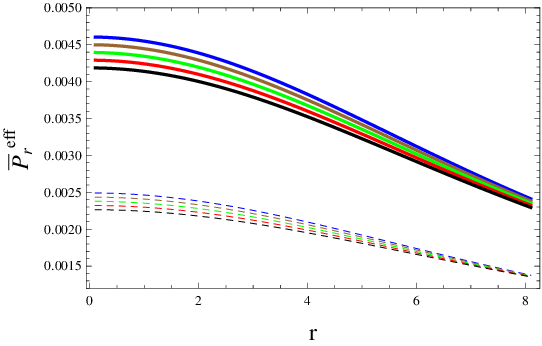,width=0.475\linewidth}
\epsfig{file=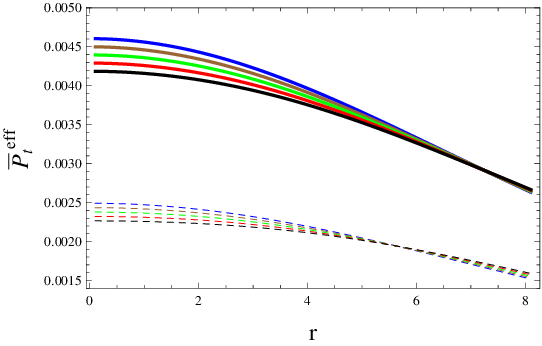,width=0.475\linewidth}
\epsfig{file=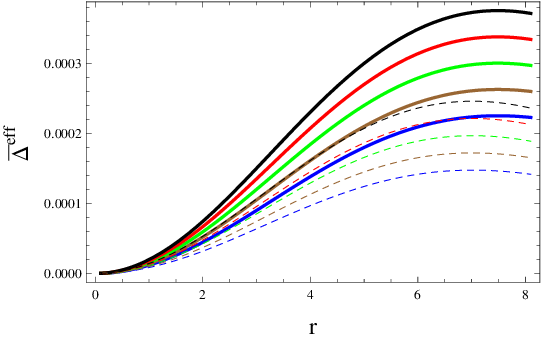,width=0.475\linewidth} \caption{Plots of
$\rho^{eff},\bar{P}_r^{eff},\bar{P}_t^{eff}$ and
$\bar{\Delta}^{eff}$ versus $r$ corresponding to $\lambda=0.3$
(solid), $0.25$ (dashed), $\delta=0.12$ (blue), $0.14$ (brown),
$0.16$ (green), $0.18$ (red), $0.2$ (black) and $Q_0=0.01$ for
solution I.}
\end{figure}
We now interpret this solution through graphical analysis of the
effective parameters $\big(\rho^{eff}$, $\bar{P}_r^{eff}$,
$\bar{P}_t^{eff}\big)$ and the anisotropy, $\bar{\Delta}^{eff}$. We
have done this analysis for the star Her X-1 with mass
$M_0=0.85M_{\bigodot}$ and radius $R=8.1km$ \cite{41}. Throughout
this analysis, we have used two values of the Rastall parameter and
charge parameter given by $\lambda=0.3,0.25$ and $Q_0=0.01,2$
respectively, with the decoupling constant as
$\delta=1.2,1.4,1.6,1.8,2~.$ The values used for these parameters
have been selected after a long trial for which they have been found
to induce the desired behavior in the obtained model. We see that
the behavior of the effective parameters (density, radial and
tangential pressures) is consistent with the behavior of compact
objects, i.e., finite, positive and maximum at the core and
monotonically decreasing towards the surface of the star. This is
shown in Figures \textbf{1} and \textbf{2}.
\begin{figure}\center
\epsfig{file=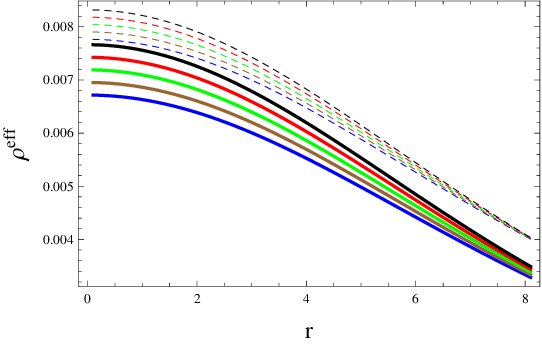,width=0.475\linewidth}
\epsfig{file=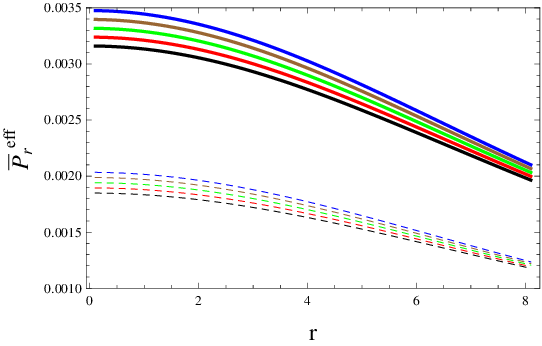,width=0.475\linewidth}
\epsfig{file=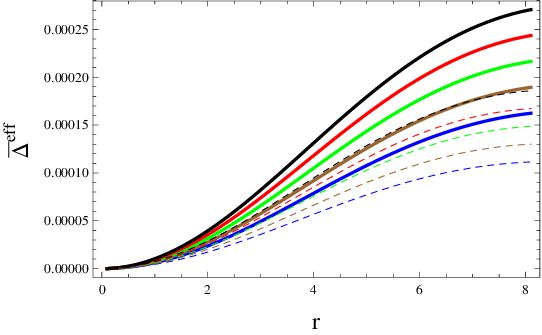,width=0.475\linewidth} \caption{Plots of
$\rho^{eff},\bar{P}_r^{eff},\bar{P}_t^{eff}$ and
$\bar{\Delta}^{eff}$ versus $r$ corresponding to $\lambda=0.3$
(solid), $0.25$ (dashed), $\delta=0.12$ (blue), $0.14$ (brown),
$0.16$ (green), $0.18$ (red), $0.2$ (black) and $Q_0=2$ for solution
I.}
\end{figure}

It can be observed from both figures that the density incurs higher
values for $\lambda=0.25$ when compared to $\lambda=0.3$. This leads
to the conclusion that a reduction in the Rastall parameter renders
a more dense interior of compact stars. The effect of the decrement
in the Rastall parameter is seen to correspond to lower values of
the radial and tangential pressures. Furthermore, the radial and
tangential pressures reach the same magnitude near the core, causing
an anisotropy that disappears at that point and increases towards
the surface. This positive anisotropy illustrates an outward
directed pressure, leading to the anti-gravitational force which
helps in the stability of compact structure. An increase in the
Rastall parameter results in more anisotropy. With regards to the
charge, however, we observe that a higher value induces a less dense
interior for compact objects. Figures \textbf{1} and \textbf{2} show
that the effective parameters as well as anisotropy are slightly
lowered by an increase in charge.

\subsection{Solution II}

We now adopt a density-like constraint for the purpose of obtaining
a new charged anisotropic solution. This constraint is given by
\begin{equation}\label{55}
\Theta_0^0(r)=\rho(r).
\end{equation}
Using Eqs.\eqref{30} and \eqref{44} in the constraint above, we have
\begin{align}\nonumber
\frac{g^{\ast\prime}}{r}+\frac{g^\ast}{r^2}&=\frac{C}{2 \sqrt{C r^2}
\left(C r^2+1\right)^2 \left(2 A+B r \sqrt{C r^2}\right)^2}\bigg[4
A^2 \sqrt{C r^2} \bigg(C (4 \lambda -1) r^2\\\nonumber &+12 \lambda
-6\bigg)+4 A B C r^3 \left(C(4 \lambda -1) r^2+12 \lambda
-6\right)+B^2 r^2 \sqrt{C r^2} \\\nonumber &\times\bigg(C r^2
\left(C(4 \lambda -1) r^2-4 \lambda -2\right)-16 \lambda +4\bigg)-2
B r \ln \bigg(A+\frac{1}{2} B r\\\nonumber &\times \sqrt{C
r^2}\bigg)\bigg(2 B (4 \lambda -1) r \sqrt{C r^2} \left(C
r^2+1\right)\log \left(A+\frac{1}{2} B r \sqrt{C
r^2}\right)\\\label{56} &+2 A \bigg(C (8 \lambda +1) r^2+12 \lambda
\bigg)+B r \sqrt{C r^2} \left(3 C r^2+4 \lambda +2\right)\bigg)
\bigg].
\end{align}
\begin{figure}\center
\epsfig{file=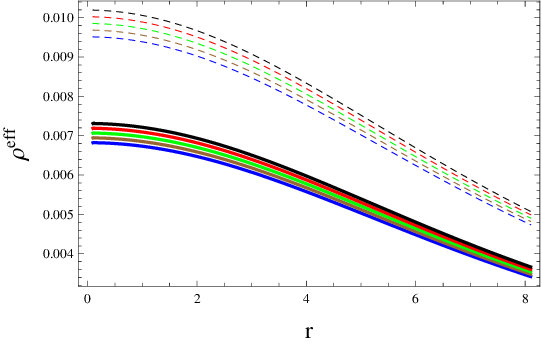,width=0.475\linewidth}
\epsfig{file=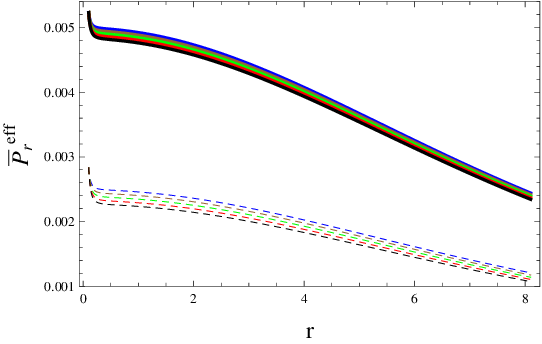,width=0.475\linewidth}
\epsfig{file=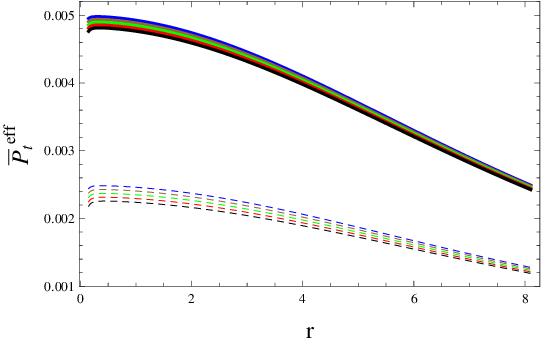,width=0.475\linewidth}
\epsfig{file=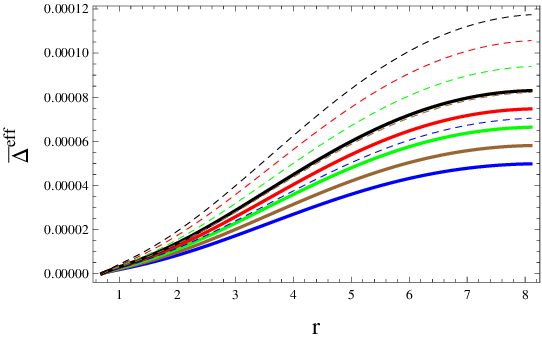,width=0.475\linewidth} \caption{Plots of
$\rho^{eff},\bar{P}_r^{eff},\bar{P}_t^{eff}$ and
$\bar{\Delta}^{eff}$ versus $r$ corresponding to $\lambda=0.3$
(solid), $0.25$ (dashed), $\delta=0.01$ (blue), $0.03$ (brown),
$0.05$ (green), $0.07$ (red), $0.09$ (black) and $Q_0=0.01$ for
solution II.}
\end{figure}
\begin{figure}\center
\epsfig{file=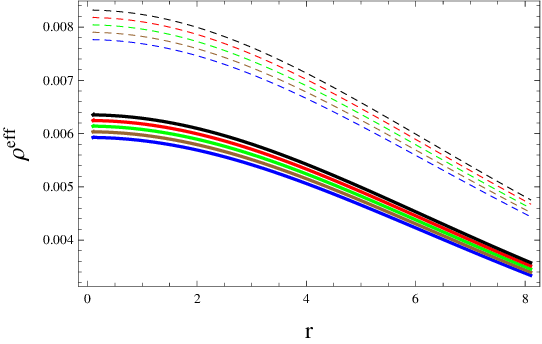,width=0.475\linewidth}
\epsfig{file=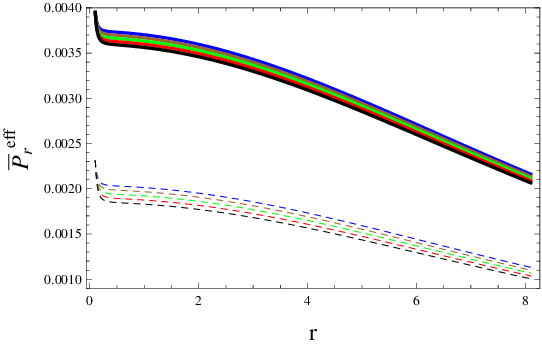,width=0.475\linewidth}
\epsfig{file=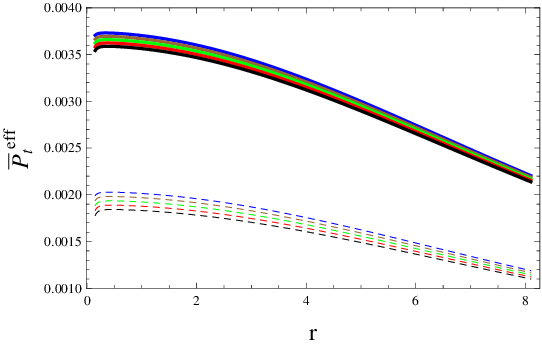,width=0.475\linewidth}
\epsfig{file=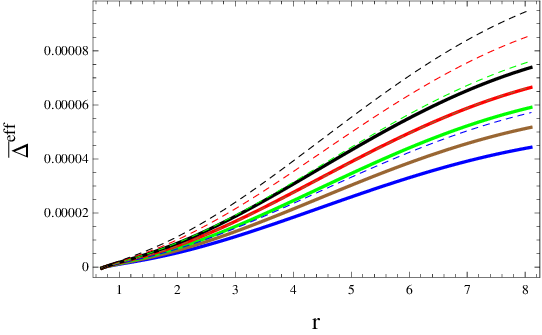,width=0.475\linewidth} \caption{Plots of
$\rho^{eff},\bar{P}_r^{eff},\bar{P}_t^{eff}$ and
$\bar{\Delta}^{eff}$ versus $r$ corresponding to $\lambda=0.3$
(solid), $0.25$ (dashed), $\delta=0.01$ (blue), $0.03$ (brown),
$0.05$ (green), $0.07$ (red), $0.09$ (black) and $Q_0=2$ for
solution II.}
\end{figure}
Owing to the absence of an exact solution to the differential
equation \eqref{56}, we obtain a numerical approximation. Figures
\textbf{3} and \textbf{4} show the plots of the effective parameters
and anisotropy for this solution. For the graphical analysis, we
have used the same values of the parameters as in solution I. It is
seen from these figures that the effective parameters are finite and
attain a positive as well as maximum value at the core whilst
monotonically decreasing towards the boundary. It is also observed
that lowering the value of the Rastall parameter enhances a denser
interior of compact stars whilst inducing lower values for the
radial and tangential pressures. The radial and tangential pressures
attain almost similar values near the core thus inducing a positive
anisotropy that vanishes near the core. With respect to the charge,
it is observed that increasing its value enhances a less dense
interior for compact structures and lower values of radial and
tangential pressures. A reduced anisotropy is also enhanced by an
increase in charge. It is worthy to highlight that the alteration of
the Rastall and charge parameters induces the same effects in both
solutions I and II.

\subsection{Physical Viability and Stability}
\begin{figure}\center
\epsfig{file=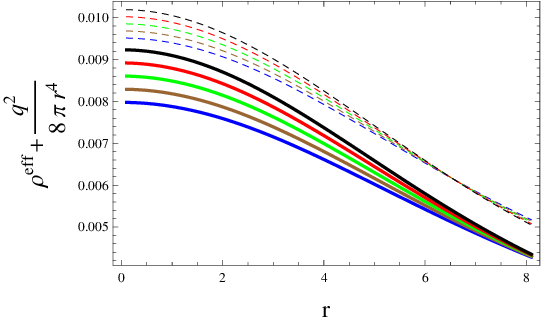,width=0.475\linewidth}
\epsfig{file=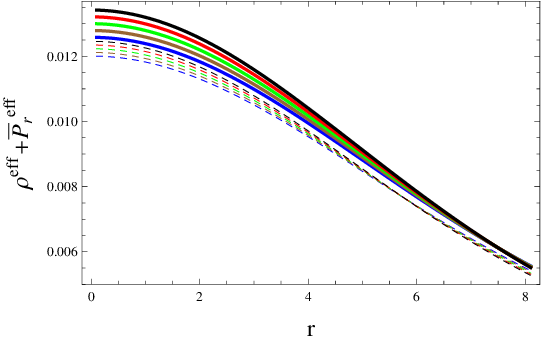,width=0.475\linewidth}
\epsfig{file=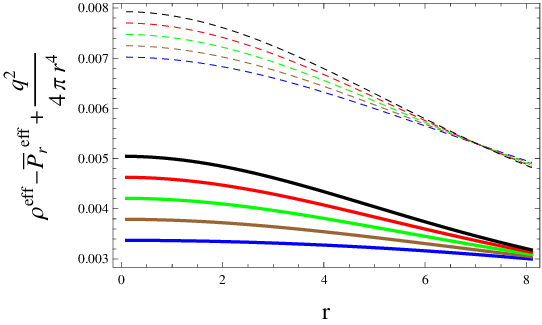,width=0.475\linewidth}
\epsfig{file=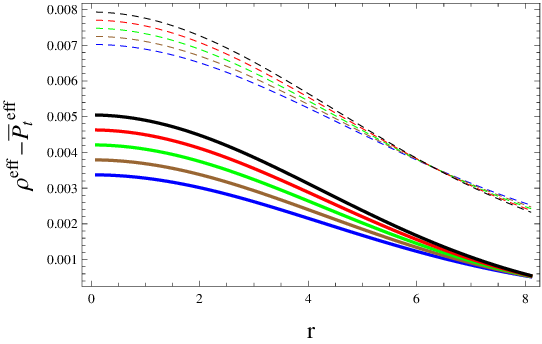,width=0.475\linewidth}
\epsfig{file=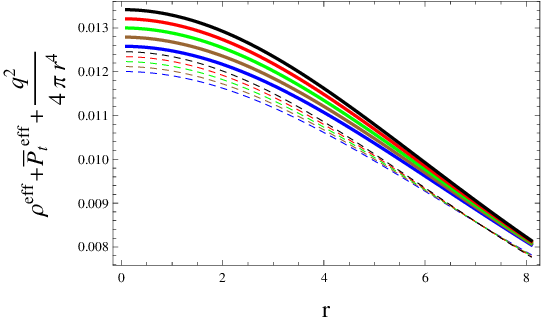,width=0.475\linewidth}
\epsfig{file=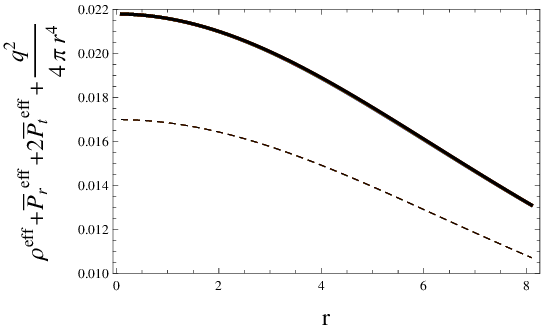,width=0.475\linewidth} \caption{Plots of energy
conditions versus $r$ corresponding to $\lambda=0.3$ (solid), $0.25$
(dashed), $\delta=0.12$ (blue), $0.14$ (brown), $0.16$ (green),
$0.18$ (red), $0.2$ (black) and $Q_0=0.01$ for solution I.}
\end{figure}

Here, we investigate physical properties such as viability and
stability of the obtained solutions. To study the physical
viability, we explore the energy conditions. Through the energy
conditions, we determine the presence of ordinary matter in the
stellar interior. These conditions are bounds on the stress-energy
tensor and are categorized as dominant, strong, weak and null energy
conditions. For charged distribution of matter in Rastall theory,
these conditions are given as
\begin{align}\nonumber
\rho^{eff}+\frac{q^2}{8\pi r^4}\geq
0,\quad\rho^{eff}+\bar{P}_r^{eff}\geq 0,\\\nonumber
\rho^{eff}-\bar{P}_r^{eff}+\frac{q^2}{4\pi r^4}\geq
0,\quad\rho^{eff}-\bar{P}_t^{eff}\geq 0,\\\nonumber
\rho^{eff}+\bar{P}_t^{eff}+\frac{q^2}{4\pi r^4}\geq
0,\quad\rho^{eff}+\bar{P}_r^{eff}+2\bar{P}_t^{eff}+\frac{q^2}{4\pi
r^4}.
\end{align}
Figures \textbf{5-8} show that these bounds are met by both
solutions thus implying their physical viability. The radial and
tangential EoS parameters given by
$\omega_r=\frac{\bar{P}_r^{eff}}{\rho^{eff}}$ and
$\omega_t=\frac{\bar{P}_t^{eff}}{\rho^{eff}},$ respectively,
comprise an important component of self-gravitating objects and can
be used to determine their physical viability. A necessary and
sufficient condition for physical viability using the EoS parameters
is that $0\leq\omega_r\leq 1$ and $0\leq\omega_t\leq 1$ \cite{42}.
These are shown in Figures \textbf{9} and \textbf{10} that depict
the physical viability of both solutions.
\begin{figure}\center
\epsfig{file=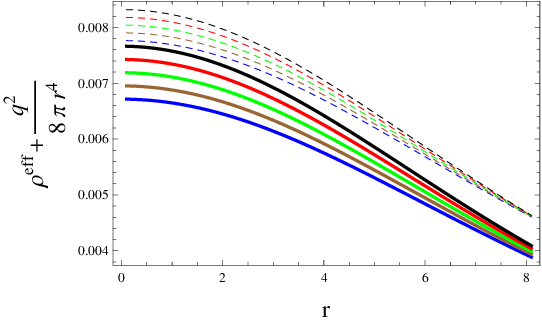,width=0.475\linewidth}
\epsfig{file=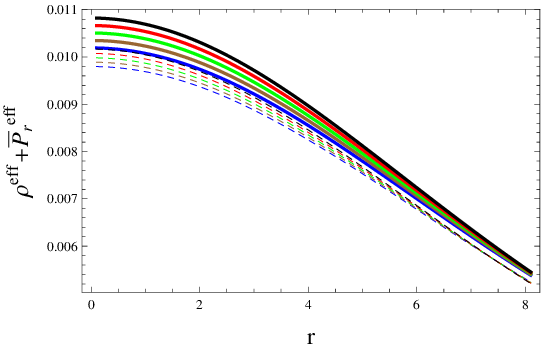,width=0.475\linewidth}
\epsfig{file=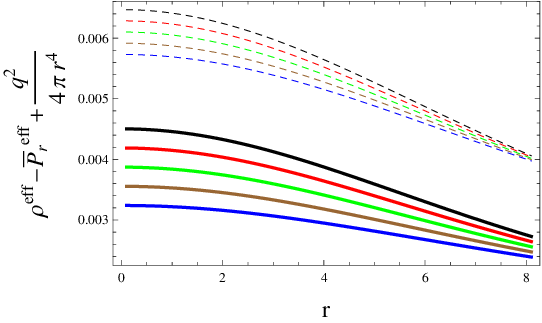,width=0.475\linewidth}
\epsfig{file=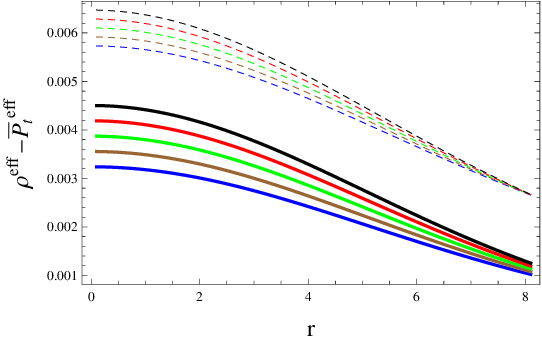,width=0.475\linewidth}
\epsfig{file=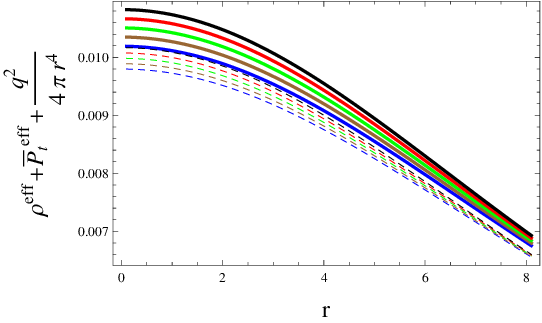,width=0.475\linewidth}
\epsfig{file=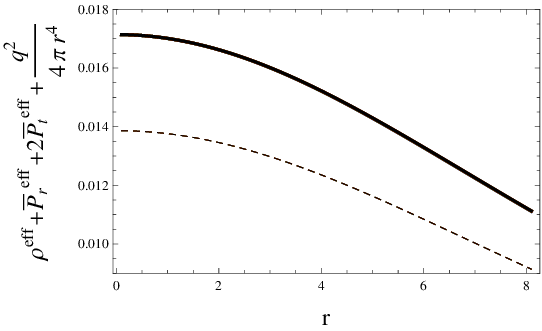,width=0.475\linewidth} \caption{Plots of energy
conditions versus $r$ corresponding to $\lambda=0.3$ (solid), $0.25$
(dashed), $\delta=0.12$ (blue), $0.14$ (brown), $0.16$ (green),
$0.18$ (red), $0.2$ (black) and $Q_0=2$ for solution I.}
\end{figure}
\begin{figure}\center
\epsfig{file=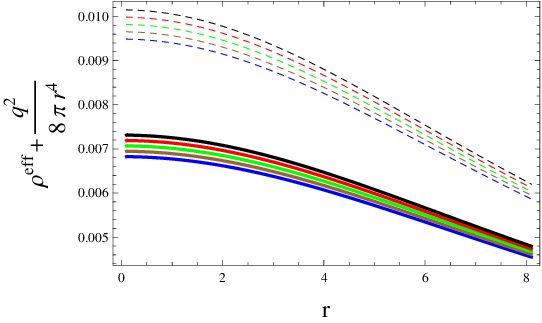,width=0.475\linewidth}
\epsfig{file=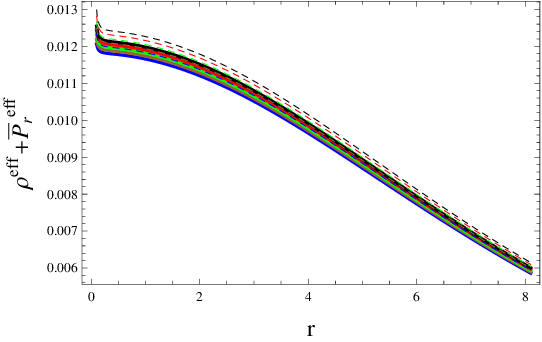,width=0.475\linewidth}
\epsfig{file=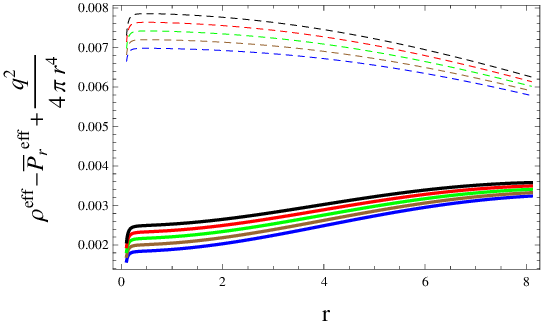,width=0.475\linewidth}
\epsfig{file=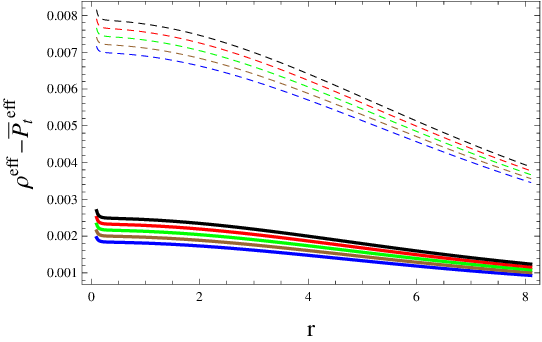,width=0.475\linewidth}
\epsfig{file=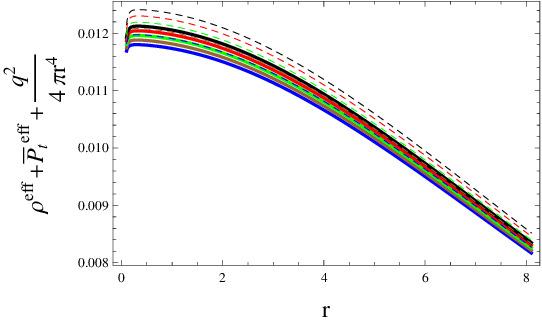,width=0.475\linewidth}
\epsfig{file=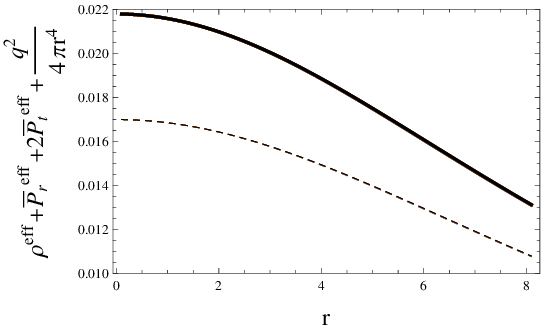,width=0.475\linewidth} \caption{Plots of energy
conditions versus $r$ corresponding to $\lambda=0.3$ (solid), $0.25$
(dashed), $\delta=0.12$ (blue), $0.14$ (brown), $0.16$ (green),
$0.18$ (red), $0.2$ (black) and $Q_0=0.01$ for solution II.}
\end{figure}
\begin{figure}\center
\epsfig{file=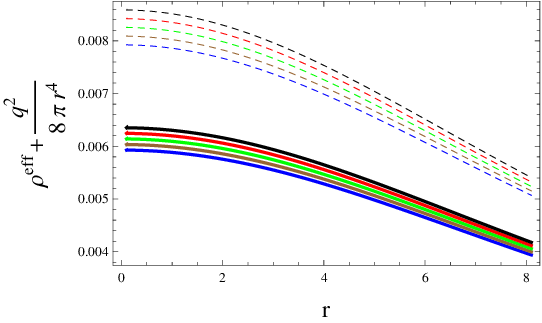,width=0.475\linewidth}
\epsfig{file=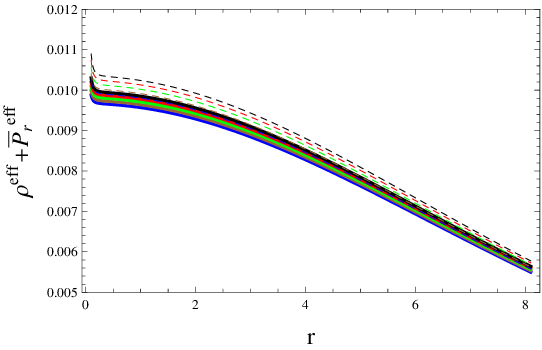,width=0.475\linewidth}
\epsfig{file=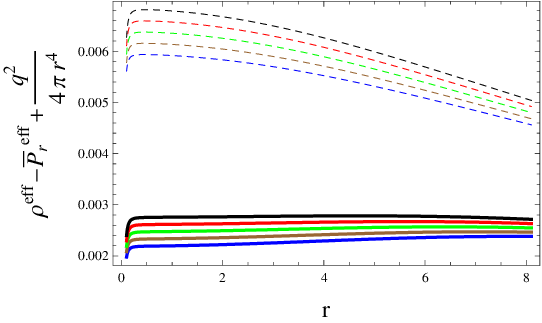,width=0.475\linewidth}
\epsfig{file=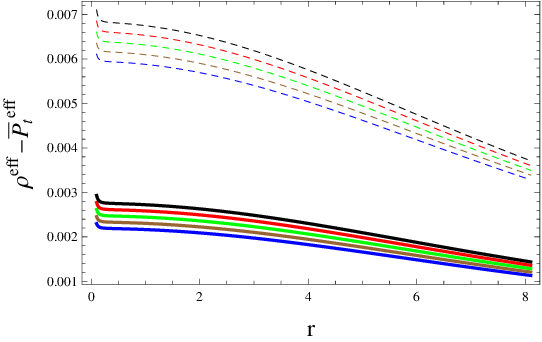,width=0.475\linewidth}
\epsfig{file=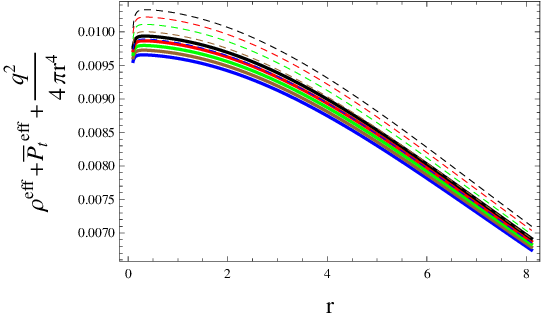,width=0.475\linewidth}
\epsfig{file=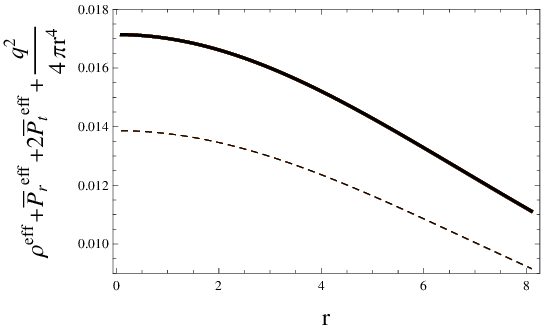,width=0.475\linewidth} \caption{Plots of energy
conditions versus $r$ corresponding to $\lambda=0.3$ (solid), $0.25$
(dashed), $\delta=0.12$ (blue), $0.14$ (brown), $0.16$ (green),
$0.18$ (red), $0.2$ (black) and $Q_0=2$ for solution II.}
\end{figure}

For a static and spherically symmetric self-gravitating body, we
determine the mass by the equation
\begin{equation}\label{57}
m(r)=4\pi\int_0^R\rho^{eff}r^2dr.
\end{equation}
With the initial condition $m(0)=0,$ the mass of anisotropic
spherical structure can be numerically evaluated from the above
equation. Using this mass, we can determine the compactness $u(r)$
and surface redshift $Z_s(r)$ defined by $u(r)=\frac{m(r)}{r}$ and
$Z_s(r)=\frac{1}{\sqrt{1-2u(r)}}-1,$ respectively. For viable
stellar configuration, Buchdahl \cite{43} determined the limit
$u(r)<\frac{4}{9}~.$ The surface redshift parameter is used to
evaluate the increment induced by the strong gravitational pull of a
celestial body on the wavelength of electromagnetic radiations. For
a perfect fluid distribution of matter, Buchdahl restricted this
parameter to the limit $Z_s(r)<2$ at the surface of the star. For an
anisotropic configuration, however, the limit becomes $Z_s(r)\leq
5.211$ \cite{44}. Figures \textbf{11} and \textbf{12} illustrate
that both solutions meet the required limits for compactness and
surface redshift.
\begin{figure}\center
\epsfig{file=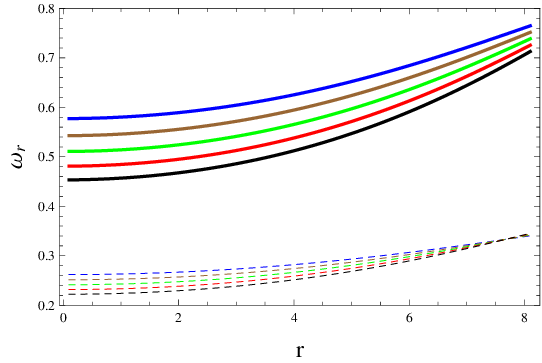,width=0.475\linewidth}
\epsfig{file=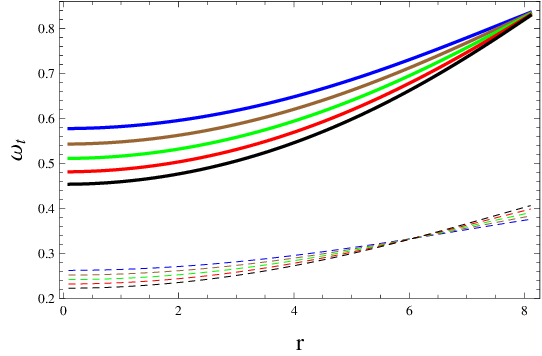,width=0.475\linewidth}
\epsfig{file=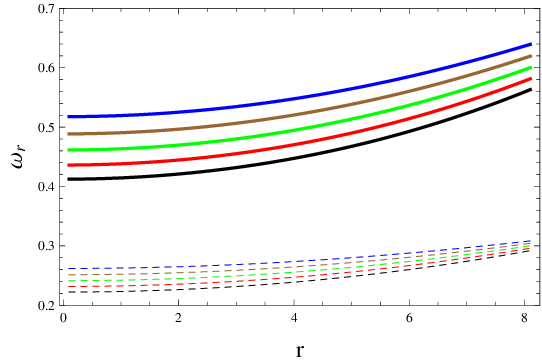,width=0.475\linewidth}
\epsfig{file=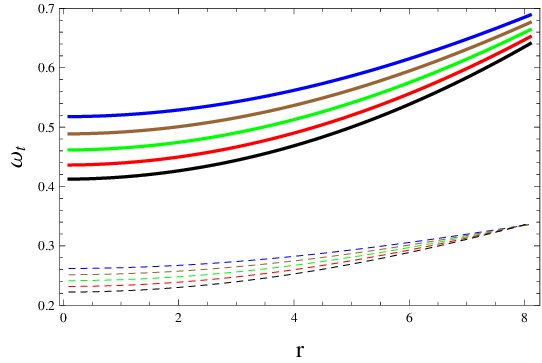,width=0.475\linewidth}\caption{Plots of
radial and tangential EoS parameters versus $r$ corresponding to
$\lambda=0.3$ (solid), $0.25$ (dashed), $\delta=0.12$ (blue), $0.14$
(brown), $0.16$ (green), $0.18$ (red), $0.2$ (black), $Q_0=0.01$
(top row) and $Q_0=2$ (bottom row) for solution I}
\end{figure}
\begin{figure}\center
\epsfig{file=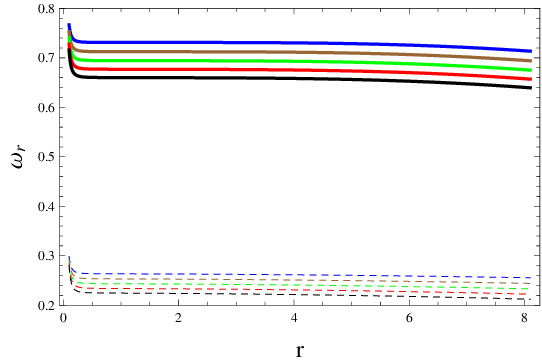,width=0.475\linewidth}
\epsfig{file=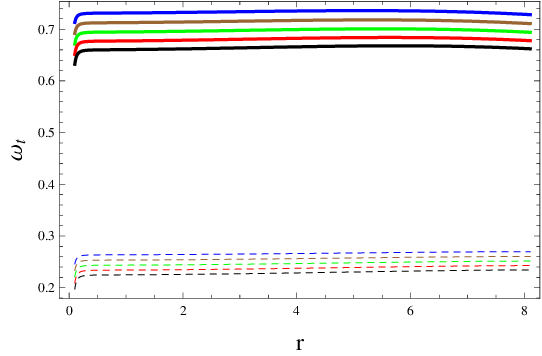,width=0.475\linewidth}
\epsfig{file=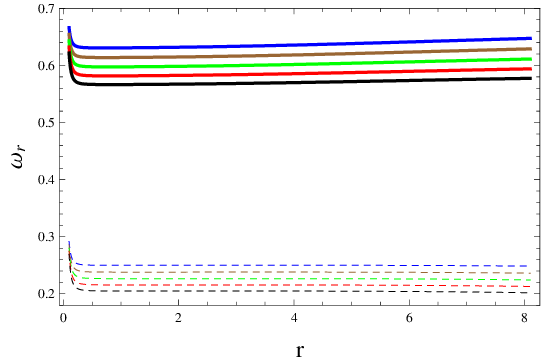,width=0.475\linewidth}
\epsfig{file=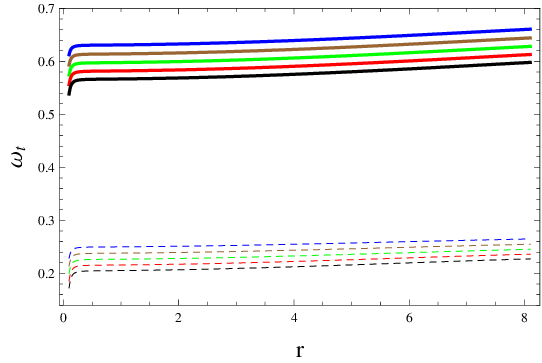,width=0.475\linewidth}\caption{Plots of
radial and tangential EoS parameters versus $r$ corresponding to
$\lambda=0.3$ (solid), $0.25$ (dashed), $\delta=0.12$ (blue), $0.14$
(brown), $0.16$ (green), $0.18$ (red), $0.2$ (black), $Q_0=0.01$
(top row) and $Q_0=2$ (bottom row) for Solution II.}
\end{figure}
\begin{figure}\center
\epsfig{file=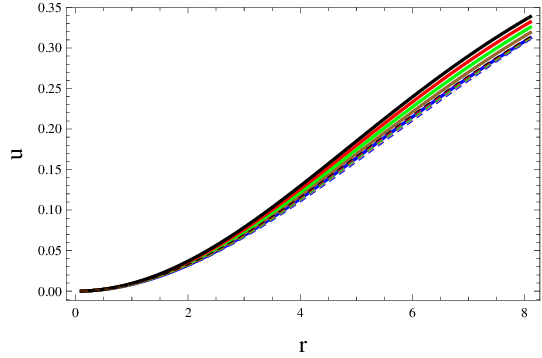,width=0.475\linewidth}
\epsfig{file=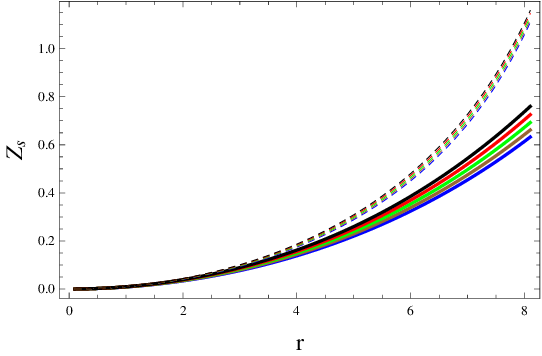,width=0.475\linewidth}
\epsfig{file=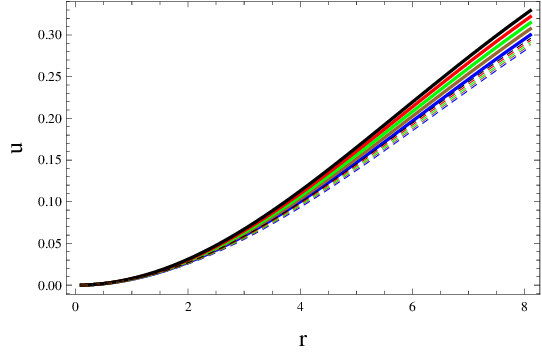,width=0.475\linewidth}
\epsfig{file=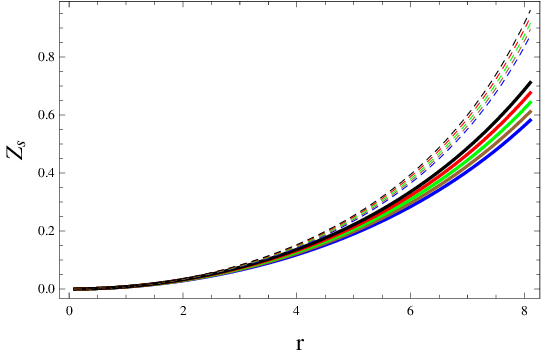,width=0.475\linewidth}\caption{Plots of
compactness and surface redshift versus $r$ corresponding to
$\lambda=0.3$ (solid), $0.25$ (dashed), $\delta=0.12$ (blue), $0.14$
(brown), $0.16$ (green), $0.18$ (red), $0.2$ (black), $Q_0=0.01$
(top row) and $Q_0=2$ (bottom row) for Solution I.}
\end{figure}
\begin{figure}\center
\epsfig{file=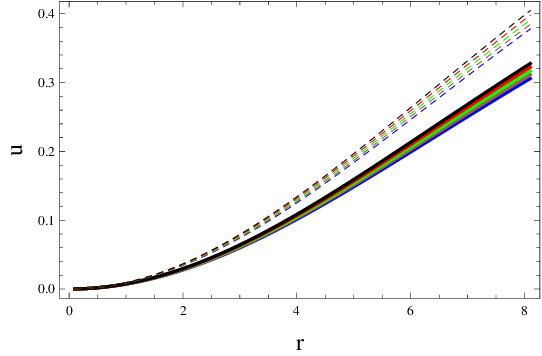,width=0.475\linewidth}
\epsfig{file=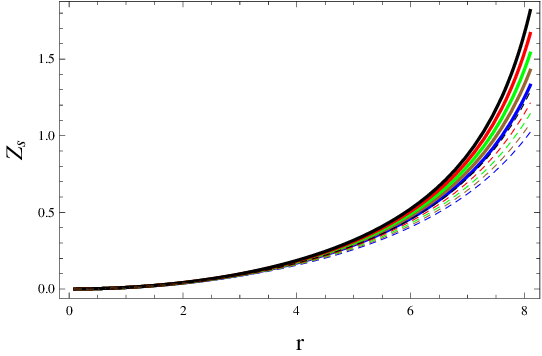,width=0.475\linewidth}
\epsfig{file=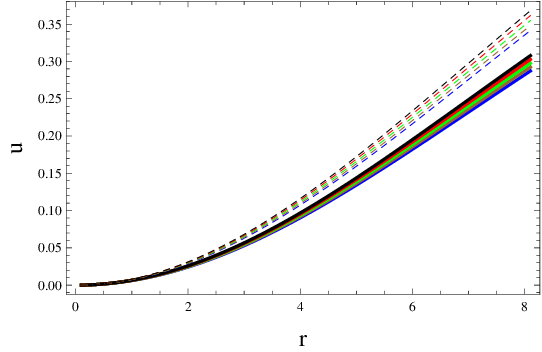,width=0.475\linewidth}
\epsfig{file=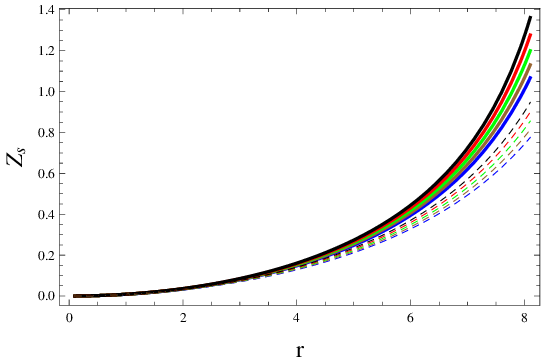,width=0.475\linewidth}\caption{Plots of
compactness and surface redshift versus $r$ corresponding to
$\lambda=0.3$ (solid), $0.25$ (dashed), $\delta=0.12$ (blue), $0.14$
(brown), $0.16$ (green), $0.18$ (red), $0.2$ (black), $Q_0=0.01$
(top row) and $Q_0=2$ (bottom row) for Solution II.}
\end{figure}
\begin{figure}\center
\epsfig{file=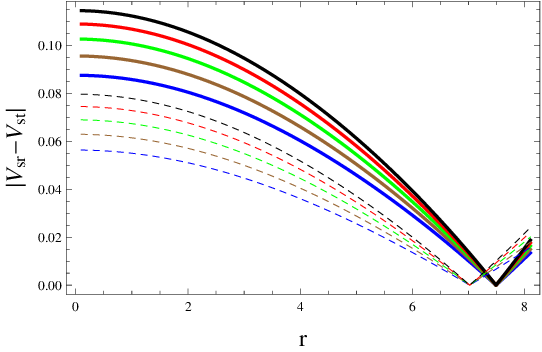,width=0.475\linewidth}
\epsfig{file=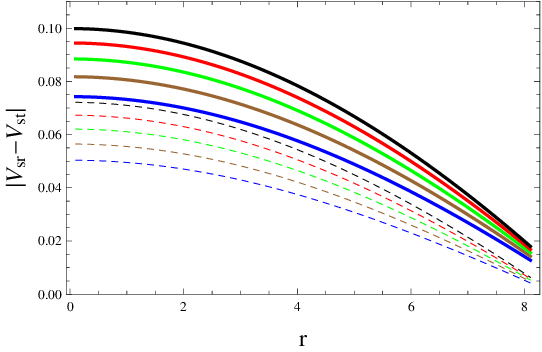,width=0.475\linewidth}\caption{Plots of
$|V_{st}^2-V_{sr}^2|$ versus $r$ corresponding to $\lambda=0.3$
(solid), $0.25$ (dashed), $\delta=0.12$ (blue), $0.14$ (brown),
$0.16$ (green), $0.18$ (red), $0.2$ (black), $Q_0=0.01$ (left) and
$Q_0=2$ (right) for Solution I.}
\end{figure}
\begin{figure}\center
\epsfig{file=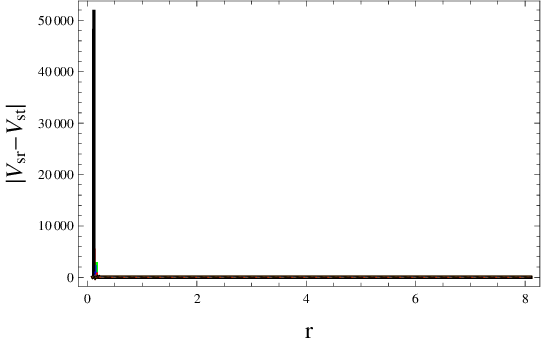,width=0.475\linewidth}
\epsfig{file=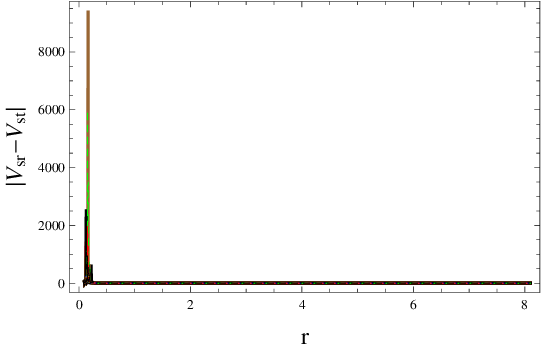,width=0.475\linewidth}\caption{Plots of
$|V_{st}^2-V_{sr}^2|$ versus $r$ corresponding to $\lambda=0.3$
(solid), $0.25$ (dashed), $\delta=0.12$ (blue), $0.14$ (brown),
$0.16$ (green), $0.18$ (red), $0.2$ (black), $Q_0=0.01$ (left) and
$Q_0=2$ (right) for Solution II.}
\end{figure}
\begin{figure}\center
\epsfig{file=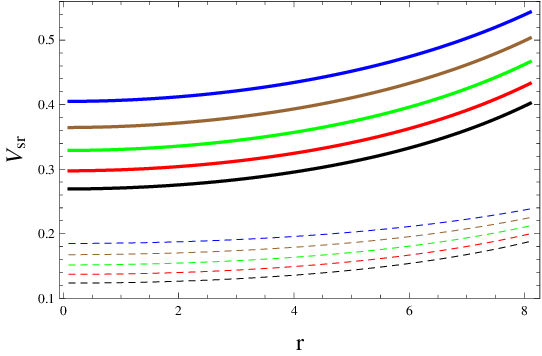,width=0.475\linewidth}
\epsfig{file=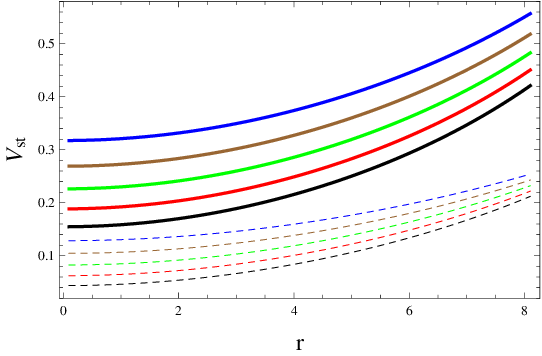,width=0.475\linewidth}
\epsfig{file=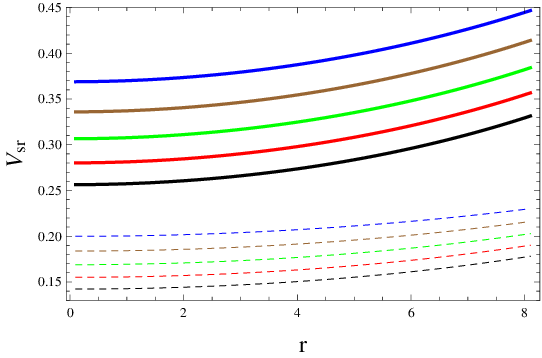,width=0.475\linewidth}
\epsfig{file=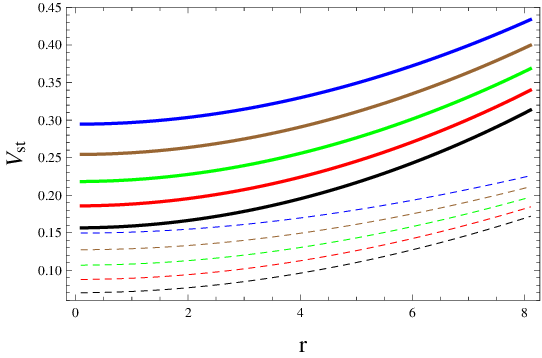,width=0.475\linewidth}\caption{Plots of radial
and tangential sound speeds versus $r$ corresponding to
$\lambda=0.3$ (solid), $0.25$ (dashed), $\delta=0.12$ (blue), $0.14$
(brown), $0.16$ (green), $0.18$ (red), $0.2$ (black), $Q_0=0.01$
(top row) and $Q_0=2$ (bottom row) for Solution I.}
\end{figure}
\begin{figure}\center
\epsfig{file=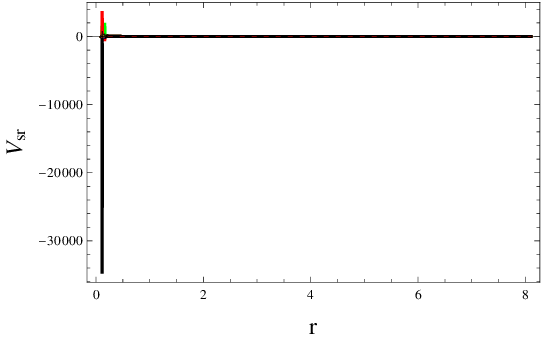,width=0.475\linewidth}
\epsfig{file=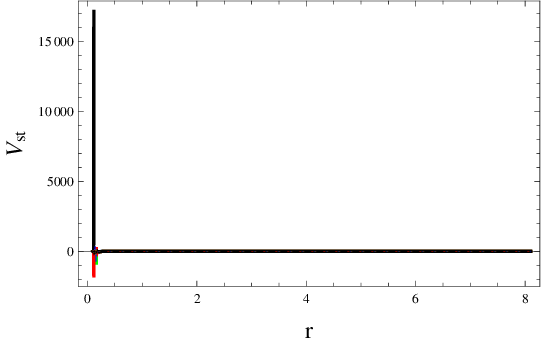,width=0.475\linewidth}
\epsfig{file=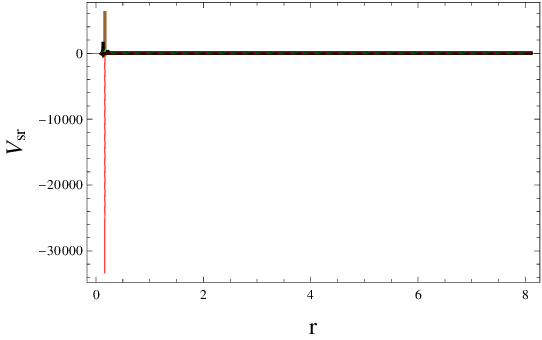,width=0.475\linewidth}
\epsfig{file=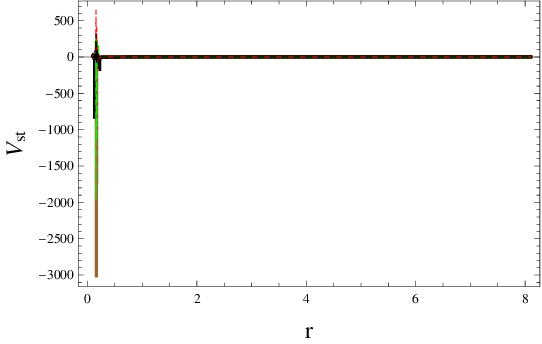,width=0.475\linewidth}\caption{Plots of radial
and tangential sound speeds versus $r$ corresponding to
$\lambda=0.3$ (solid), $0.25$ (dashed), $\delta=0.12$ (blue), $0.14$
(brown), $0.16$ (green), $0.18$ (red), $0.2$ (black), $Q_0=0.01$
(top row) and $Q_0=2$ (bottom row) for Solution II.}
\end{figure}

Finally, to complete the physical analysis, we discuss the stability
of both solutions. For this purpose, we use both the Herrera
cracking approach \cite{38} as well as the causality conditions.
With the Herrera cracking technique, the stability of compact
objects is implied if $0\leq |V_{st}^2-V_{sr}^2|\leq 1$, where
$V_{st}^2=\frac{d\bar{P}_t^{eff}}{d\rho^{eff}}$ and
$V_{sr}^2=\frac{d\bar{P}_r^{eff}}{d\rho^{eff}}$ denote the
tangential and radial sound speeds, respectively. Figures
\textbf{13} and \textbf{14} show that solution I is stable for both
values of the charge parameter while solution II is unstable. To
further verify this result, the causality condition is employed.
Here, the condition for stability is that the squared sound speed
must lie in the closed interval $\left[0,~1\right]$, i.e., $0\leq
V_s^2\leq 1$, otherwise stated, $0\leq V_{sr}^2\leq 1$ and $0\leq
V_{st}^2\leq 1$. It turns out that solution I is stable while
solution II is unstable as shown by the plots in Figures \textbf{15}
and \textbf{16}.

\section{Conclusions}

In this paper, we have used gravitational decoupling through the MGD
approach to explore the effects of electromagnetic field on
anisotropic spherical solutions in Rastall gravity. Using this
approach, the known isotropic Finch-Skea ansatz has been extended to
include the effects of charge in anisotropic domain. To achieve
this, we have added a new source $\Theta_{\gamma\varrho}$ to the
effective charged isotropic stress-energy tensor
$\bar{T}^{(eff)}_{\gamma\varrho}$, leading to the effective field
equations with anisotropic distribution of matter. By deforming only
the radial metric component of the known solution, the effective
field equations \eqref{18}-\eqref{20} are decoupled into two sets.
The first set \eqref{26}-\eqref{28} corresponds to the Rastall field
equations for isotropic matter configuration while the second set
\eqref{30}-\eqref{32} corresponds to the added source
$\Theta_{\gamma\varrho}$. The matching conditions at the stellar
surface have also been studied in detail, for an outer RN spacetime.

For the Rastall parameter $\lambda=0.3,0.25$ with the decoupling
constant $~\delta=0.12~,0.14~,0.16~,0.18~,0.2~$ and charge
$Q_0=0.01,2,$ the physical behavior of the effective parameters
$\big(\rho^{eff},\bar{P}_r^{eff},\bar{P}_t^{eff}\big)$ and
anisotropic pressure $\big(\bar{\Delta}^{eff}\big)$ for both
solutions have been examined. It is found that reducing the Rastall
and charge parameters induces a more dense stellar interior in both
solutions. However, an increment in charge coincides with a reduced
compactness of stellar interior in both solutions. A rise in the
decoupling parameter $\delta$ is also observed to induce a more
dense interior as well as an increased anisotropy in both solutions.
The generated anisotropy in both cases has been found to be
positive, implying an outward directed pressure which produces the
anti-gravitational force required to keep the compact object in an
equilibrium state.

The physical viability of both solutions, which we have tested
through graphical analysis of the energy conditions, shows that
these solutions are physically acceptable. Finally, we have
discussed the stability of both solutions through two different
criteria. We have found that solution I turns to be stable while
solution II is unstable. It is worth mentioning here that two
anisotropic solutions corresponding to a pressure-like and
density-like constraint were found in GR \cite{30}. However, only
the solution corresponding to the pressure-like constraint was found
to be viable and stable. Here, we have found that both solutions are
viable but only the first solution is stable. We can thus conclude
that more viable solutions are obtained in Rastall theory than GR.\\\\
\section*{Appendix}
\begin{align}\nonumber
\rho^{eff}&=\frac{C}{2 \sqrt{C r^2} \left(C r^2+1\right)^3 \left(2
A+B r \sqrt{C r^2}\right)^2} \bigg[-\left(C r^2+1\right)\\\nonumber
&\times \bigg(4 A^2 \sqrt{C r^2} \left(C (4 \lambda -1) r^2+12
\lambda -6\right)-2 B r \ln \left(A+\frac{1}{2} B r \sqrt{C
r^2}\right)\\\nonumber &\times \bigg(2 B (4 \lambda -1) r \sqrt{C
r^2} \left(C r^2+1\right) \ln \left(A+\frac{1}{2} B r \sqrt{C
r^2}\right)\\\nonumber &+2 A \left(C (8 \lambda +1) r^2+12 \lambda
\right)+B r \sqrt{C r^2} \left(3 C r^2+4 \lambda +2\right)\bigg)+4 A
B C r^3\\\nonumber &\times \left(C (4 \lambda -1) r^2+12 \lambda
-6\right)+B^2 r^2 \sqrt{C r^2} \bigg(C r^2 \left(C (4 \lambda -1)
r^2-4 \lambda -2\right)\\\nonumber &-16 \lambda
+4\bigg)\bigg)+\frac{2Cr^2\delta}{\left(B C r^3 \left(4 \ln
\left(A+\frac{1}{2} B r \sqrt{C r^2}\right)+1\right)+2 A \sqrt{C
r^2}\right)^2}\\\nonumber &\times 16 A^4 \sqrt{C r^2} \left(C^2 (2
\lambda -1) r^4+4 C (\lambda -1) r^2+18 \lambda -3\right)+32 A^3 B C
r^3\\\nonumber &\times \left(C^2 (2 \lambda -1) r^4+4 C (\lambda -1)
r^2+18 \lambda -3\right)-8 A^2 B^2 r^2 \sqrt{C r^2}\\\nonumber
&\times \left(C r^2 \left(C^2 (2 \lambda -1) r^4+40 C \lambda r^2+22
\lambda -3\right)+32 \lambda -4\right)+2 B r\\\nonumber &\times \ln
\left(A+\frac{1}{2} B r \sqrt{C r^2}\right) \bigg(B r \ln
\left(A+\frac{1}{2} B r \sqrt{C r^2}\right)\\\nonumber &\times
\bigg(4 A^2 \sqrt{C r^2} \left(C r^2 \left(C (4 \lambda -11) r^2-8
(\lambda +1)\right)-44 \lambda +3\right)+4 A B C r^3\\\nonumber
&\times \left(C r^2 \left(C (20 \lambda -9) r^2+40 \lambda
-4\right)-12 \lambda +5\right)-4 B C (4 \lambda +1) r^3\\\nonumber
&\times \left(C r^2+1\right) \left(2 A \left(C r^2+3\right)+B r
\sqrt{C r^2} \left(1-C r^2\right)\right) \ln \left(A+\frac{1}{2} B r
\sqrt{C r^2}\right)\\\nonumber &-B^2 r^2 \left(C r^2\right)^{3/2}
\left(C r^2 \left(C (28 \lambda +15) r^2+8 (5 \lambda +2)\right)+44
\lambda +1\right)\bigg)\\\nonumber &+2 \bigg(8 A^3 \left(C r^2
\left(C r^2 \left(C (1-2 \lambda ) r^2-14 \lambda +1\right)-\lambda
+3\right)-9 \lambda +3\right)\\\nonumber &-4 A^2 B r \sqrt{C r^2}
\left(C r^2 \left(C r^2 \left(C (2 \lambda -1) r^2+16 \lambda
+3\right)-35 \lambda -3\right)+11 \lambda -7\right)\\\nonumber &+2 A
B^2 C r^4 \left(C r^2 \left(C r^2 \left(C (2 \lambda -1) r^2-14
\lambda -11\right)+9 \lambda -7\right)-35 \lambda
+3\right)\\\nonumber &+B^3 r^3 \left(C r^2\right)^{3/2} \left(C r^2
\left(C r^2 \left(C (2 \lambda -1) r^2+4 \lambda -7\right)+5 \lambda
-7\right)-17 \lambda -1\right)\bigg)\bigg)\\\nonumber &-8 A B^3 Cr^5
\left(C r^2 \left(C r^2 \left(3 C (2 \lambda -1) r^2+36 \lambda
-8\right)+34 \lambda -9\right)+20 \lambda -4\right)\\\nonumber &+B^4
r^4 \left(Cr^2\right)^{3/2} \bigg(C r^2 \left(7 C^2 (1-2 \lambda )
r^4+4 C (3 \lambda +5) r^2+90 \lambda +21\right)\\\nonumber &+48
\lambda +8\bigg) \bigg], \\\nonumber \bar{P}_r^{eff}&=\frac{C}{2
\sqrt{C r^2} \left(C r^2+1\right)^2 \left(2 A+B r \sqrt{C
r^2}\right)^2} \bigg[-2 \delta \bigg(4 A^2 \sqrt{C r^2} \bigg(C (2
\lambda -1) r^2\\\nonumber &+6 \lambda -1\bigg)-2 B^2 (4 \lambda +1)
r^2 \sqrt{C r^2} \left(C r^2+1\right) 2\ln\left(A+\frac{1}{2} B r
\sqrt{C r^2}\right)\\\nonumber &+4 B r \left(A \left(2 C (1-2
\lambda ) r^2-6 \lambda +2\right)+B r \sqrt{C r^2} \left(C
r^2-\lambda +1\right)\right)\\\nonumber
&\times\ln\left(A+\frac{1}{2} B r \sqrt{C r^2}\right)+4 A B C r^3
\left(C (2 \lambda -1) r^2+6 \lambda -1\right)\\\nonumber &+B^2 r^2
\sqrt{C r^2} \left(C^2 (2 \lambda -1) r^4-C (2 \lambda +1) r^2-8
\lambda \right)\bigg)+4 A^2\\\nonumber &\times \sqrt{C r^2} \left(c
(4 \lambda -1) r^2+12 \lambda -2\right)+2 B r \ln\left(A+\frac{1}{2}
B r \sqrt{C r^2}\right)\\\nonumber &\times \bigg(-2 B (4 \lambda -1)
r \sqrt{C r^2} \left(C r^2+1\right) \ln\left(A+\frac{1}{2} B r
\sqrt{C r^2}\right)\\\nonumber &+A \left(2 C (3-8 \lambda ) r^2-24
\lambda +8\right)+B r \sqrt{C r^2} \left(C r^2-4 \lambda
+2\right)\bigg)\\\nonumber &+4 A B C r^3 \left(C (4 \lambda -1)
r^2+12 \lambda -2\right)+B^2 r^2 \sqrt{C r^2} \bigg(C r^2 \bigg(C (4
\lambda -1) r^2\\\nonumber &-4 \lambda +2\bigg)-16 \lambda
+4\bigg)\bigg],
\\\nonumber \bar{P}_t^{eff}&=\frac{C}{2 \sqrt{C r^2} \left(C
r^2+1\right)^3 \left(2 A+B r \sqrt{C r^2}\right)^3}\\\nonumber
&\times\bigg[\frac{-2Cr^2\delta}{\left(4 B C r^3 \ln
\left(A+\frac{1}{2} B r \sqrt{C r^2}\right)+2 A \sqrt{C r^2}+B C
r^3\right)^2}\\\nonumber &\times\bigg(-2 A B^4 r^4 \left(C^2 (10
\lambda -7) r^4+C (6 \lambda -19) r^2+36 \lambda
-12\right)\\\nonumber &\times \left(C r^2\right)^{3/2}-32 A^5
\left(C (2 \lambda +1) r^2-6 \lambda +1\right) \sqrt{C r^2}-16 A^3
B^2 r^2\\\nonumber &\times \left(3 C^2 (6 \lambda +1) r^4+C (1-8
\lambda ) r^2+14 \lambda -2\right) \sqrt{C r^2}-32 B^5 C^2
r^9\\\nonumber &\times \left(C r^2+1\right)^2 (4 \lambda +1) 5\ln
\left(A+\frac{1}{2} B r \sqrt{C r^2}\right)+8 B^4 C r^6 \left(C
r^2+1\right)\\\nonumber &\times \bigg(B C \left(C (4 \lambda +9)
r^2-12 \lambda +7\right) r^3+2 A \sqrt{C r^2} \bigg(3 C (1-12
\lambda ) r^2\\\nonumber &-52 \lambda +1\bigg)\bigg)4\ln
\left(A+\frac{1}{2} B r \sqrt{C r^2}\right)+4 B^3 C r^5 \bigg(B^2 C
\bigg(4 C^3 (2 \lambda -1) r^6\\\nonumber &-3 C^2 (12 \lambda +7)
r^4-C (112 \lambda +25) r^2-84 \lambda -8\bigg) r^4+4 A B \sqrt{C
r^2} \bigg(4 C^3\\\nonumber &\times (2 \lambda -1) r^6+4 C^2r^4(11
\lambda -1)+C (32 \lambda +9) r^2-20 \lambda +9\bigg) r+4
A^2\\\nonumber &\times \bigg(4 C^3 (2 \lambda -1) r^6+C^2r^4(1-4
\lambda )+C (19-80 \lambda ) r^2-84 \lambda
+14\bigg)\bigg)\\\nonumber &\times 3\ln\left(A+\frac{1}{2} B r
\sqrt{C r^2}\right)+2 B^2 r^2 \bigg(2 A B^2 r^2 \bigg(10 C^3 (2
\lambda -1) r^6-7 C^2\\\nonumber &\times (12 \lambda +5) r^4+C
(5-188 \lambda ) r^2-252 \lambda +30\bigg) \left(C
r^2\right)^{3/2}+8 A^3 \bigg(6 C^3r^6\\\nonumber &\times(2 \lambda
-1)+C^2 (36 \lambda -17) r^4+Cr^2(36 \lambda -1)-44 \lambda
+10\bigg) \sqrt{C r^2}\\\nonumber &+B^3 C^2 r^7 \bigg(2 C^3 (2
\lambda -1) r^6+C^2 (1-20 \lambda ) r^4+C (23-64 \lambda ) r^2-96
\lambda\\\nonumber &+20\bigg)+4 A^2 B C r^3 \bigg(14 C^3 (2 \lambda
-1) r^6+C^2 (132 \lambda -53) r^4+-40 \lambda +20\\\nonumber &+C
(232 \lambda -19) r^2\bigg)\bigg) 2\ln\left(A+\frac{1}{2} B r
\sqrt{C r^2}\right)-80 A^4 B C r^3\bigg((2 \lambda +1)\\\nonumber
&\times Cr^2-6 \lambda +1\bigg)+B^5 C^2 r^9 \left(3 C^2 (2 \lambda
+1) r^4+c (10 \lambda +7) r^2-4 \lambda +4\right)\\\nonumber &-8 A^2
B^3 C r^5 \left(C^2 (22 \lambda -1) r^4+C (12 \lambda -7) r^2+30
\lambda -6\right)-2 B r \bigg(-4 A\\\nonumber &\times B^3 r^3\left(4
C^3 (2 \lambda -1) r^6+C^2 (4 \lambda -7) r^4+C (8-3 \lambda )
r^2-55 \lambda +11\right) \left(C r^2\right)^{3/2}\\\nonumber &+16
A^3 B r \left(C^2 (8 \lambda +7) r^4+C (4-43 \lambda ) r^2+5 \lambda
-3\right) \sqrt{C r^2}+B^4 C^2 r^8 \bigg(4 C^3r^6\\\nonumber
&\times(1-2 \lambda )+C^2 (3-10 \lambda ) r^4+11 C (2 \lambda -1)
r^2+52 \lambda -10\bigg)+16 A^4 \bigg(3 C^2r^4\\\nonumber &\times(2
\lambda +1)+Cr^2(1-16 \lambda )+6 \lambda -2\bigg)-8 A^2 B^2 C r^4
\bigg(2 C^3 (2 \lambda -1) r^6-11 C^2\\\nonumber &\times(2 \lambda
+1) r^4+Cr^2(11 \lambda -1)-47 \lambda +8\bigg)\bigg) \ln
\left(A+\frac{1}{2} B r \sqrt{c r^2}\right)\bigg)\\\nonumber
&+\left(C r^2+1\right) \left(2 A+B r \sqrt{C r^2}\right)\bigg(4 A^2
\sqrt{C r^2} \left(C (4 \lambda -1) r^2+12 \lambda -2\right)+2 B
r\\\nonumber &\times \ln\left(A+\frac{1}{2} B r \sqrt{C r^2}\right)
\bigg(-2 B r \sqrt{C r^2}(4 \lambda -1) \left(C r^2+1\right) \ln
\left(A+\frac{1}{2} B r \sqrt{C r^2}\right)\\\nonumber &+A \left(2 C
(3-8 \lambda ) r^2-24 \lambda +8\right)+B r \sqrt{C r^2} \left(C
r^2-4 \lambda +2\right)\bigg)+4 A B C r^3 \bigg(12
\lambda\\\nonumber & -2+Cr^2(4 \lambda -1)\bigg)+B^2 r^2 \sqrt{C
r^2} \bigg(C r^2\left(C (4 \lambda -1) r^2-4 \lambda +2\right)-16
\lambda +4\bigg)\bigg)\bigg],
\\\nonumber \bar{\Delta}^{eff}&=\frac{-\delta  \left(C
r^2\right)^{3/2}}{\left(C r^2+1\right)^3 \left(B C r^3 \left(4 \ln
\left(A+\frac{1}{2} B r \sqrt{C r^2}\right)+1\right)+2 A \sqrt{C
r^2}\right)^2}\\\nonumber &\times\frac{1}{\left(2 A+B r \sqrt{C
r^2}\right)^3} \bigg[2 A B^4 r^2 \bigg(C r^2 \bigg(C \left(5 C (1-2
\lambda ) r^2-26 \lambda +17\right) r^2\\\nonumber &+12 (\lambda
+2)\bigg)-12 (\lambda -1)\bigg) \left(C r^2\right)^{3/2}+32 A^5 C
\bigg(C (1-2 \lambda ) r^2-10 \lambda\\\nonumber & +1\bigg) \sqrt{C
r^2}-16 A^3 B^2 \bigg(C \left(C \left(5 C (2 \lambda -1) r^2+54
\lambda -7\right) r^2+14 \lambda -4\right) r^2\\\nonumber &+10
\lambda -2\bigg) \sqrt{C r^2}-80 A^4 B C^2 r^3 \left(c (2 \lambda
-1) r^2+10 \lambda -1\right)+B^5 C^2 r^7\\\nonumber &\times \left(C
\left(C \left(C (1-2 \lambda ) r^2+6 \lambda +5\right) r^2+20
\lambda +8\right) r^2+4 (\lambda +1)\right)-8 A^2 B^3\\\nonumber
&\times C r^3 \left(C \left(C\left(5 C (2 \lambda -1) r^2+50 \lambda
-11\right) r^2+6 (3 \lambda -2)\right) r^2+18 \lambda
-6\right)\\\nonumber &+2 B \ln \left(A+\frac{1}{2} B r \sqrt{C
r^2}\right) \bigg(-4 A B^3 r^2 \bigg(C \bigg(C \left(4 C (2 \lambda
-1) r^2+26 \lambda -5\right)
\\\nonumber &\times r^2-21 \lambda -8\bigg) r^2+17 \lambda -7\bigg)
\left(Cr^2\right)^{3/2}-16 A^3 B \bigg(C \bigg(C \bigg(8 C(2 \lambda
-1) r^2\\\nonumber &+66 \lambda -5\bigg) r^2-11 \lambda +4\bigg)
r^2-5 \lambda +1\bigg) \sqrt{C r^2}-16 A^4 C r \bigg(C \bigg(4 C (2
\lambda -1) r^2\\\nonumber &+34 \lambda -3\bigg) r^2-2 \lambda
+1\bigg)+B^4 C^2 r^7 \bigg(C \bigg(C (10 \lambda +3) r^2+20 \lambda
+11\bigg) r^2\\\nonumber &-18 \lambda +8\bigg)-8 A^2 B^2 C r^3
\bigg(C \left(C \left(10 C (2 \lambda -1) r^2+96 \lambda -7\right)
r^2+11 \lambda +1\right) r^2\\\nonumber &+19 \lambda -2\bigg)+B \ln
\left(A+\frac{1}{2} B r \sqrt{C r^2}\right) \bigg(-2 A B^2 r^2
\bigg(C \bigg(2 C \bigg(7 C (2 \lambda -1) r^2\\\nonumber &+84
\lambda +16\bigg) r^2+68 \lambda +61\bigg) r^2+96 \lambda +15\bigg)
\left(C r^2\right)^{3/2}-8 A^3 \bigg(C \bigg(2
C\bigg(Cr^2\\\nonumber &\times(2 \lambda -1)-4 \lambda +8\bigg)
r^2-76 \lambda +23\bigg) r^2-8 \lambda +5\bigg) \sqrt{C r^2}+B^3 C^2
r^7\\\nonumber &\times \left(C \left(2 C \left(3 C (1-2 \lambda )
r^2-8 \lambda +1\right) r^2+40 \lambda +1\right) r^2-12 \lambda
+5\right)-4 A^2 B C r^3\\\nonumber &\times \left(C \left(2 C \left(5
C (2 \lambda -1) r^2-8 \lambda +25\right) r^2-288 \lambda +85\right)
r^2-84 \lambda +25\right)+2 B C r^3\\\nonumber &\times \ln
\left(A+\frac{1}{2} B r \sqrt{C r^2}\right) \bigg(B^2 C \bigg(C r^2
\bigg(C \left(4 C (2 \lambda -1) r^2-20 \lambda -33\right) r^2-64
\lambda\\\nonumber & -49\bigg)-4 (13 \lambda +5)\bigg) r^4+4 A B
\sqrt{C r^2} \bigg(C \bigg(4 C \left(C (2 \lambda -1) r^2+19 \lambda
-4\right) r^2\\\nonumber &+112 \lambda -15\bigg) r^2+28 \lambda
-3\bigg) r+2 B \left(C r^2+1\right) \ln \left(A+\frac{1}{2} B r
\sqrt{C r^2}\right)\\\nonumber &\times \bigg(B C \left(c (20 \lambda
+13) r^2+4 \lambda +11\right) r^3-4 B C \left(C r^2+1\right) (4
\lambda +1)\\\nonumber &\times \ln \left(A+\frac{1}{2} B r \sqrt{C
r^2}\right) r^3+2 A \sqrt{C r^2} \left(C (7-20 \lambda ) r^2-36
\lambda +5\right)\bigg) r\\\nonumber&+4 A^2 \left(C \left(C\left(4
C(2 \lambda -1) r^2+44 \lambda -11\right) r^2+32 \lambda -5\right)
r^2-20 \lambda +2\right)\bigg)\bigg)\bigg)\bigg].
\end{align}\\\\
\textbf{Data Availability:} No data was used for the research
described in this paper.

\end{document}